\RequirePackage{fix-cm}
\documentclass[smallextended]{svjour3}       
\usepackage{algorithm}
\usepackage[noend]{algpseudocode}
\smartqed  
\usepackage{graphicx}
\usepackage{epstopdf}
\usepackage{subfigure}
\usepackage{amsmath}
\usepackage{cite}
\begin{document}


\title{Adaptive Blind CDMA Receivers Based on ICA Filtered Structures}
\vspace{2ex}
\author{Zaid Albataineh         \and
        Fathi M. Sale} 

\maketitle




\begin{abstract}
Code Division Multiple Access (CDMA) is a channel access method adopted by various radio technologies world-wide. In particular, CDMA is used as an access method in many mobile standards such as CDMA2000 and WCDMA.  We address the problem of blind multiuser equalization in the wideband CDMA systems in the noisy multipath propagation environment. Herein, we propose three new blind receiver schemes based on variations of Independent Component Analysis (ICA) within several filtering structures. These adaptive blind CDMA (ABC) receivers do not require knowledge of the propagation parameters or spreading code sequences of the users\textemdash they primarily exploit the natural assumption of statistical independence among the symbol signals. We also develop three semi-blind adaptive detectors by incorporating new adaptive methods into the standard Rake receiver structure. Extensive comparative case-studies, based on Bit error rate (BER) performance are carried out for as a function of (i) the number of users, (ii) the number of symbols per user, and (iii) the signal to noise ratio (SNR).The conventional detectors include the baseline Linear Minimum mean squared error (LMMSE) detector. The results show that the proposed methods outperform other detectors in estimating the symbol signals from the received mixed CDMA signals. Moreover, the new blind detectors mitigate the multi access interference (MAI) in CDMA.
\end{abstract}

\section{Introduction}
\label{intro}
Code Division Multiple Access (CDMA) is a channel access method ubiquitously used in various modalities and platforms worldwide. It is based on spread-spectrum technology as is found, e.g., in third-generation (3G) cellular telephony, terrestrial and satellite communications systems, and indoor wireless networks [1-2], [9]. Although, LTE (4G) is utilized by several cellular companies inside and outside the U.S., their networks are still not fully built, and LTE coverage is still not universal. Thus, most of the older 2G and 3G systems are ubiquitous and exist in parallel with the newer 4G systems worldwide. In the U.S., companies like AT\&T and T-Mobile use GSM/WCDMA/HSPA while Verizon, Sprint, and MetroPCS use CDMA2000/EV-DO [3-5]. Moreover, the newer LTE wireless interface is incompatible with the 2G and 3G networks, so that it must be operated on a separate wireless spectrum. While 4G technology is intended to eventually replace the 3G technologies, it is now evident that it will take some time before LTE coverage is fully developed and widely adopted even in the developed countries [26-27]. 

As with any radio communication system, CDMA-based systems also suffer from various types of interferences. Specifically, they suffer from (i) an internal multiple access interference (MAI) due to the non-ideal cross-correlations among the users’ spreading sequences, (ii) narrow-band inter-symbol interference (ISI), and (iii) background noise at the receiver. These drawbacks affect the performance of a CDMA system.  The conventional detectors most frequently utilized to counteract CDMA interference is based on second order statistics. In highly loaded systems, conventional detectors are not considered a suitable choice. Most of the conventional detectors suffer from external interference sources and treat all interferences as a lumped background noise. In CDMA-based systems, however, the primary source of interference is MAI. This has motivated the development of numerous interference rejection techniques to overcome the MAI and the near-far problem in conventional receivers [1, 7]. Several state-of-the-art approaches have been proposed in the literature to overcome this challenge, e.g., using pilot signals and training [40].

In CDMA-based systems, multiuser detection is desirable in order to enhance channel capacity and mitigate MAI [10, 21]. Multiuser detection has been introduced to obtain an optimum multiuser detector for multi-Gaussian channels in [1]. Several suboptimal detectors have also been proposed in [6-8], to overcome the computational complexity in realizing optimal detectors. In [1] and [32-36], training pilot sequence techniques have been used to present suboptimal detectors, namely an adaptive linear detector and a zero-forcing detector. 

In [20, 21], Wang and Poor proposed the blind minimum mean square error (MMSE) and the blind de-correlating detectors. The suboptimal detector based on the linear minimum mean square error (LMMSE) method has been described in [32].In [31-36], adaptive blind detectors were proposed based on incorporating the minimum output energy with constrained optimization methods. Several subspace approaches were proposed in the literature, e.g., in [20], [23], [36]. In [10], several types of group-blind linear detectors were proposed in order to enhance the performance for the uplink and downlink channels. The key idea of these detectors is to take advantage of the cross-correlation matrix which was constructed by exploiting the correlation between successive samples of received signals.  These detectors, however, are too complex to be practically implemented, especially at the mobile unit. Also, they require information regarding signal timing and the spreading codes of the desired user. 

The aforementioned techniques periodically require the base station to send a training sequence that must be known by the mobile receiver in order to enable the latter in estimating the parameters of the channel propagation model. These parameters also attempt to capture the multiple reflections of the radio waves on encountered obstacles, i.e. buildings, cars, trees, etc. Furthermore, according to [42], it has been reported that 20\% of the bandwidth in GSM, and up to 40\% in UMTS CDMA, is devoted to the training sequence. In spite of the good performance of the training sequence techniques, the cost tends to be significantly large in terms of bandwidth. Adaptive signal processing techniques, on the other hand, provide more efficient methods for CDMA systems in the presence of high dynamic conditions as a result of the receiver mobility, the short channel codes and the fortuitous channel access. In particular, the desire to ensure a high communication rate has made blind adaptive techniques a hot topic, driven by their potential to eliminate/reduce training sessions. Moreover, blind techniques help recover symbol signals in other situations e.g., i) eavesdropping, where using the training sequence is not available, and ii) tracking, when the receiver fails to keep the desired user locked in track. It is also noted that the underlying user symbol sequences are reasonably assumed to be statistically independent. Therefore, statistical independence, or near independence, is a key assumption that makes a CDMA system suitable for the blind techniques, e.g., using information maximization [1] or minimum mutual information [6]. In [6-8], typical CDMA based systems are represented by wide stationary slowly fading multipath environment and are expressed by a linear multi-channel convolution model. Thus, the received signals in a CDMA mobile can be considered as signals generated by the linear convolutive model of statistically independent components of independent users as shown in [6], [10], [31-36]. The adaptive LMMSE detector has been originally proposed to overcome the necessary complex matrix inversion operation [38]; however, it still requires the spreading codes of all users. While the LMMSE detector maybe suitable for the uplink to the base station, as computational resources are usually abundant, it is less practical in the downlink to the receiver as computational resources are scarcer. 

This paper aims at recovering the source symbol sequences from the linear convolutive received mixture without any knowledge of the user short channelizing codes and in the absence of explicit channel identification. In essence, the paper proposes improved blind adaptive detections, based on the state space approach[37, 6],using the natural gradient method for multipath channels of CDMA-based systems. Three update-laws are derived for various filtering structures [6] and then three adaptive blind CDMA detectors are introduced for more effective MAI, ISI suppression and symbol estimation. The second contribution of the paper is three semi-blind adaptive stochastic gradient algorithms fused into the conventional Rake receiver. Specifically, we fuse algorithms based on, respectively, FastICA, RobustICA, and principle component analysis (PCA). Furthermore, higher order statistics (HOS) are exploited in order to make the proposed methods robust and secure against incomplete cross-correlation and the near-far problem in conventional detectors [42].
Extensive Monte Carlo simulations have been carried out to verify and evaluate the effectiveness of the proposed methods in estimating the users’ symbols. In summary, we provide metric comparisons in the bit error-rate (BER) as a function of (i) the number of users, and (ii) the number of symbols per user, and (iii) the signal-to-noise (SNR). The   comparisons include the proposed methods with existing and conventional ones in terms of BER performance and computational complexity.

We now set the notation used throughout the paper. Lower case letters denote scalars, bold lower case letters denote vectors, and bold upper case letters denote matrices. Moreover, the following symbols are used:
\begin{list}{*}{}
\item ${(.)^T}$ refers to the transpose operator;
\item 	${(.)^H}$ refers to the Hermitian transpose operator;
\item $trace\left( . \right)$  refers to the trace operator; 
\item $j = \sqrt { - 1} $  refers to the imaginary symbol; 
\item $diag\left( . \right)$ refers to the standard diagonal of a matrix;
\item $Diag\left( . \right)$ refers to the diagonal of a block matrix, where elements may be block matrices themselves;
\item $sgn\left( . \right)$ refers to the sign operator;
\item $E[.]$ refers to the statistical expectation operator.
\end{list}

The remainder of the paper is organized as follows. In Section II, brief descriptions and derivations of synchronous CDMA signal models in multi-path fading are presented. The conventional Rake receiver model is described in Section III. Section IV is dedicated to the derivation of adaptive update laws and to the proposed new detection schemes. The comparative simulations with summary results and conclusions are given in Section V and Section VI, respectively.

\section{CDMA SIGNAL MODEL}
\label{sec:1}
We now briefly present two signal models for a CDMA based system using one layer of channel spreading codes. Specifically, we describe the DS-CDMA signal and WCDMA signal models in a typical synchronous CDMA system usually employed, e.g., for cellphones, indoor ATM, and certain ad hoc wireless networks [1], [3].

\subsection{A DS-CDMA Receiver Signal Model}
\label{sec:1}
In a DS-CDMA system, several users share the medium simultaneously by using unique individualized code signatures. We refer to Fig. 1 below for a typical system schematic block diagram. In this paper, we assume the data transmission to be quaternary
phase shift keying (QPSK). At the mobile unit receiver, assume a total of K active users in an L multipath environment and M transmitted symbols during the observation frame time. The simplest downlink received signal model $r(t)$,  at the time sample  t over a single symbol interval   is given by  [21]

\begin{equation}
r\left( t \right) = \mathop \sum \limits_{m = 1}^M \mathop \sum \limits_{k = 1}^K \mathop \sum \limits_{l = 0}^{L-1 }{\alpha _{lm}}{b_{k,m}}{s_k}\left( {t - m{T_b} - {d_l}{T_c}} \right) + n\left( t \right)
\end{equation}
where 
\begin{list}{-}{}
\item l,k,m are path, user and symbol indices, respectively.
\item 	${\alpha _{lm}}$ is the path gain -- in the downlink model the path gain is assumed to be the same among users because all users' signals are transmitted together. Thus, the path gain ${\alpha _{lm}}$ and propagation delay factor ${d_l}$ do not depend on the user k.
\item ${b_{k,m}}$  is the kth user m symbol.
\item ${s_k}\left( . \right)$ is the kth user spreading code (chip sequence).
\item ${d_l}$ is the propagation delay factor.
\item $t,{T_b},{T_c}$ are time, symbol, and chip duration, respectively.
\item $n\left( t \right)$ is the channel additive white Gaussian noise (AWGN) with zero mean and covariance equals q.
\end{list}

The system is assumed to be time-invariant, over a small duration, which means that the channel parameters are much slower than the frequency of transmitted symbol data. Let us assume that G is the number of chips per symbol, K is the number of users, and L is the number of paths. Thus, the scalar  form of Equation (1) can be transformed to a vector form [6, 21] as:

\begin{figure}
 \includegraphics{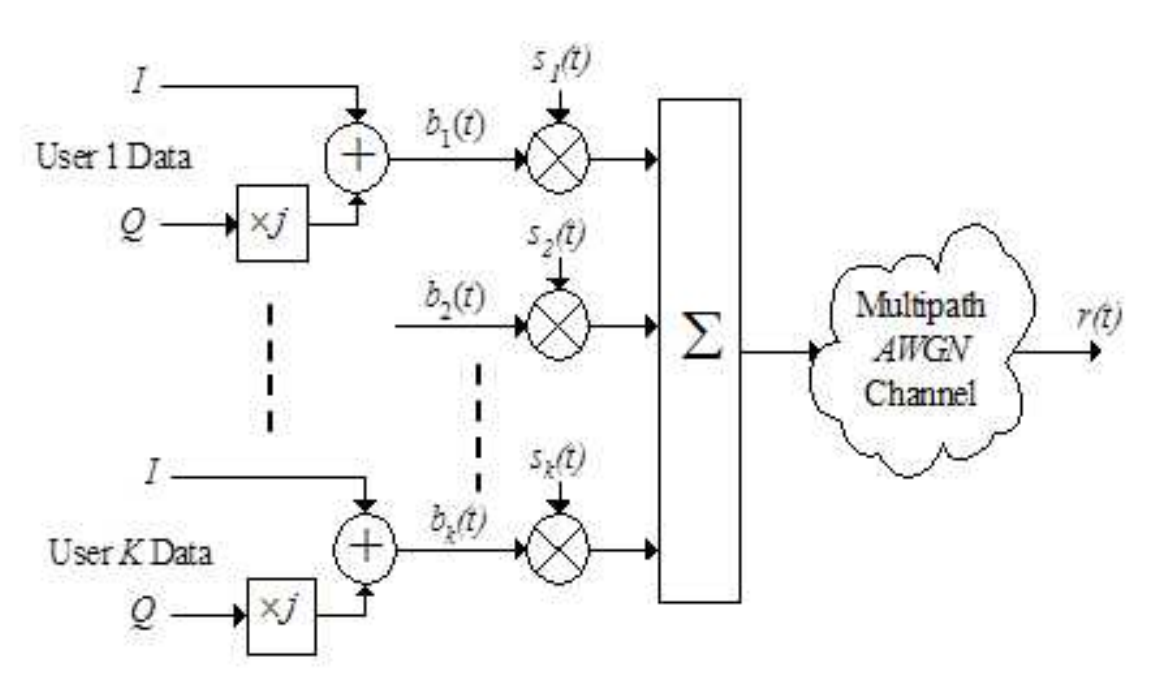}
\caption{Signal generation model for a typical QPSK DS-CDMA system}
\label{fig:1}       
\end{figure} 

\begin{equation}
\bf{ r} = \bf{H}\bf{S}\bf{ b} + \bf{ n}
\end{equation}
where ${\bf{ r}}$ is a received (G1) dimensional vector signal; ${\bf{ H}}$ is a ${\left( {{\rm{G}}1} \right){\rm{\;x\;G}}}$ matrix with ${{\rm{\;G}}1 \ge {\rm{G}} + {\rm{L}} - 1}$, which represents the multipath propagation coefficients; ${\bf{ S}}$ is a ${\rm{G\;x\;K}}$ block-diagonal matrix; ${\bf{ b}}$ is a K dimensional vector, which represents the users data symbols; and ${\bf{ n}}$ is the ${(G1)}$ dimensional channel noise vector with covariance matrix, say, ${\bf{ Q}}$. This standardized model of received signals has been used in deriving the conventional detectors, e.g., Match filter, Rake filter, blind LMMSE and other blind detectors [21]. We shall use it in our development as well. In addition, an alternative two-tap symbol signal model is given by [10]:
\begin{equation}
{\bf{ r}_n} = {\bf{ H}_0}{\bf{ b}_n} + {\bf{ H}_1}{\bf{ b}_{n - 1}} + {\bf{ n}_n} = \bf {\bar{ H}}\bf{\bar{ b}_n} + {\bf{ n}_n}
\end{equation}
where
\begin{list}{*}{}
\item ${\bf{ r}_n}$ is the total received user's signal vector; 
\item ${{\bf{H}}_0} = \left[ {{{\bf{h}}_1},\; \ldots ,\;{{\bf{h}}_{\bf{k}}}} \right]$ is the signature matrix of the current symbol vectors of all users including MAI, specifically,
\begin{equation}
{\bf{ h}_k} = \left[ {\begin{array}{*{20}{c}}
0\\
{{\bf h_k}\left( 0 \right)}\\
.\\
.\\
.\\
{{\bf h_k}\left( {G - {D_l} - 1} \right)}
\end{array}} \right]		   
 \end{equation}

\item ${\bf{ H}_1} = \left[ {\bf {{\bar{ h}_1}} ,\; \ldots ,\;\bf {{\bar{ h}_k}} } \right]$ is the signature matrix of the previous symbol vectors of all users including ISI, where, 
\begin{equation}
{\overline{\bf{ h}}_k} = \left[ {\begin{array}{*{20}{c}}
{{\bf h_k}\left( {G - {D_l}} \right)}\\
.\\
.\\
.\\
{{\bf h_k}\left( {G + L - 1} \right)}\\
0
\end{array}} \right]
 \end{equation}
${D_{l}} \in \{ 0,1, \ldots ,{{G - 1}}\}$ is the delay in chip
periods.
\item $\bf{\bar{ H}}= \left[ {\;{\bf{ H}_0}\;{\bf{ H}_1}} \right]$ is the signature matrix of all users;
\item ${\bf{ b}_n} = {\left[ {{b_1}\left( n \right),\; \ldots ,\;{b_K}\left( n \right)} \right]^T}$ are the current symbols of all users;
\item ${\bf{ b}_{n - 1}} = {\left[ {{b_1}\left( {n - 1} \right),\; \ldots ,\;{b_K}\left( {n - 1} \right)} \right]^T}$ are the previous symbols of all users;
\item ${\bf{\bar{ b}_n}} = {\left[ {\bf{ b}_n^T\;,\;\;\bf{ b}_{n - 1}^T} \right]^T}$ are the augmented two-tap symbols of all users;
\item ${\bf{ n}_n} = {\left[ {{\bf n}\left( {nG} \right), \ldots ,\; {\bf n}\left( {nG + G - 1} \right)} \right]^T}$ is the independent white composite Gaussian Noise vector. We defer further details to [6, 10].
\end{list}

In the asynchronous uplink CDMA systems, one can assume that the columns of ${\bf{ H}_0}$ and ${\bf{ H}_1}$ are mutually independent. Therefore, ${\bf{\bar{ H}}}$ is a full rank  matrix. Whereas for the synchronous downlink CDMA communication, ${\bf{\bar{ H}}}$ is full-rank with some restrictions. The main focus in this paper is on the synchronous downlink CDMA communication system, although our proposed algorithms work well in the uplink asynchronous CDMA systems [10], [30].

\subsection{WCDMA Receiver Signal Model}
\label{sec:1}

One difference between a WCDMA system and a DS-CDMA system is the presence of scrambling codes. The main cause of the MAI in WCDMA systems is the intra-cell multiple user signals sharing the same multipath channels. Fig. 2 depicts a block diagram that shows the additional code scrambling before transmission through the air interface. In Fig. 2, DPDCH stands for Dedicated Physical Data CHannel which is a term adopted in UMTS (Universal Mobile Telecommunications Systems) and a S/P block stands for serial to parallel Converter. Consequently, the basic received signal model r(t) is given by [6]:
\begin{equation}
r\left( t \right) = \mathop \sum \limits_{m = 1}^M \mathop \sum \limits_{k = 1}^K \mathop \sum \limits_{l = 0}^L {\alpha _{lm}}{b_{k,m}}{c_k}\left( {t - \;{d_l}{T_c}} \right){s_k}\left( {t - m{T_b} - {d_l}{T_c}} \right) + n\left( t \right)
 \end{equation}
where, in addition to the previous parameters, one adds ${{\rm{c}}_{\rm{k}}}\left( {\rm{t}} \right) \in \left\{ { \pm 1{\rm{\;}} \pm {\rm{j}}} \right\}$,  the complex cell-specific scrambling sequences. The remaining variables are defined in model (1). The received signal at the mobile unit is passed through a chip-matched filter and sampled at the chip rate. The received discrete vector $\bf{ r}$ in this case can be expressed as [6, 10, 21].

\begin{equation}
\bf{ r} = \bf{H}\bf{C}\bf{S}\bf{ b} + \bf{ n}
\end{equation}
where $\bf{C}$ is the $\rm{GxG}$ complex diagonal scrambling matrix with  ${\bf{C}}{{\bf{C}}^{\rm{H}}} = {{\rm{\bf{I}}}_{{\rm{GxG}}}}$ and the remaining variables are defined  similarly as in (2). The form of $\bf{C}$ is given by:

\begin{equation}
\bf{C}\; = \;\rm{diag}{\left( {{\bf{c}_1}\;\;\;{\bf{c}_2} \ldots \;\;\;\;{\bf{c}_G}\;} \right)}
\end{equation}
where
${c_i} \in \left\{ { \pm 1\; \pm j} \right\}\;\;\;\;\;\;\forall \;\;1 \le i\; \le G$

\begin{figure}
 \includegraphics{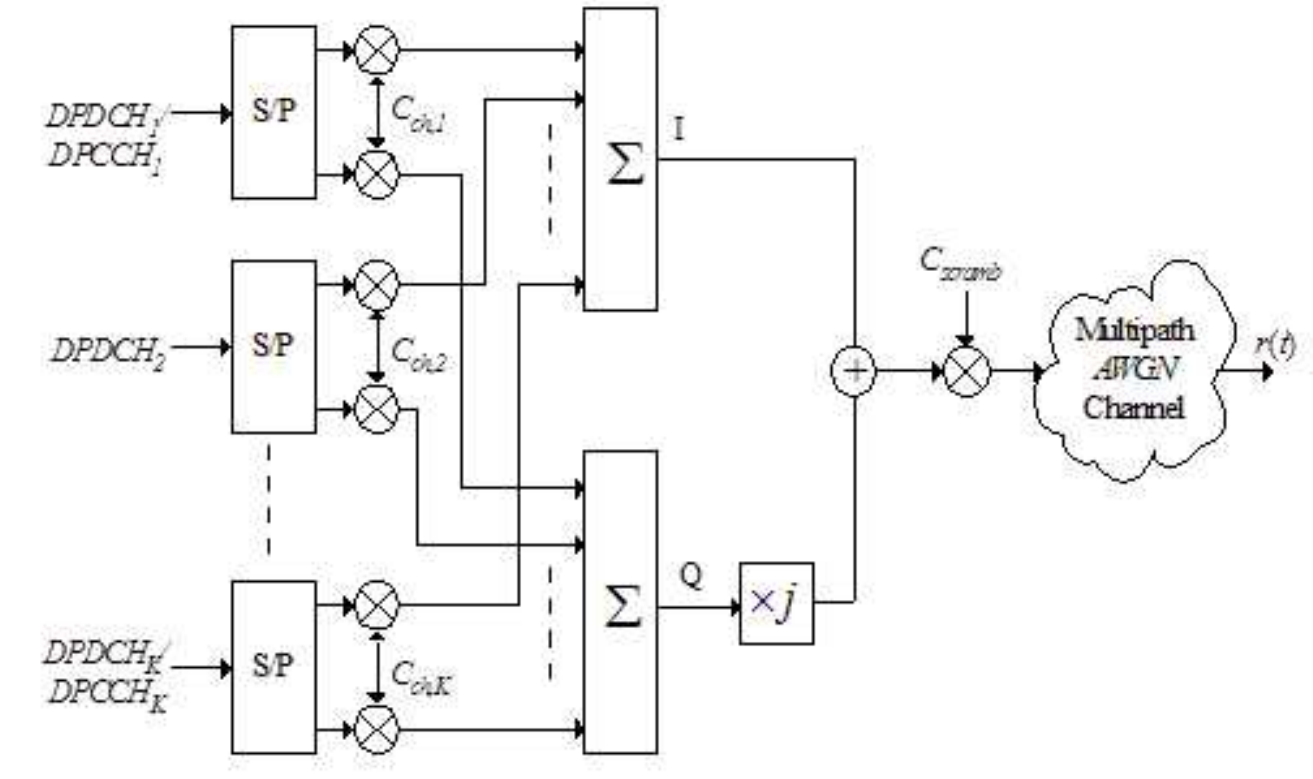}
\caption{Signal generation based on the proposed 3GPP UMTS FDD standard}
\label{fig:1}       
\end{figure} 

\section{CONVENTIONAL BLIND LINEAR MULTIUSER DETECTORS}
\label{sec:1}
We briefly describe the baseline conventional linear multiuser detectors such as the Match Filter (MF), the Rake receiver and the LMMSE detector in multipath environments. For further details, see [1, 9].
'
\subsection{Single user detector (SUD)}
\label{sec:1}

The SUD is a standard MF detector which exploits the user’s code signature to provide an estimate of the user’s symbol sequence from the received data. This detector completely ignores the presence of MAI due to other users. One can express the MF detector for the ith user in the DS-CDMA system as follows:
\begin{equation}
\bf{b}_{i,MF}^D = \bf{S}_i^H\bf{r}
\end{equation}
where ${{\bf{S}}_{\rm{i}}} = {\rm{Diag}}\left( {{{{\bf{\bar s}}}_{\rm{i}}},{{{\bf{\bar s}}}_{\rm{i}}}, \ldots ,{\rm{\;}}{{{\bf{\bar s}}}_{\rm{i}}}} \right)$ , ${{\bf{\bar s}}_{\rm{i}}} = \left[ {0{\rm{\;}}0 \ldots {\rm{\;}}{{\rm{s}}_{\rm{i}}} \ldots 0} \right]$. ${{\rm{s}}_{\rm{i}}}$ is the ith user’s signature code, $\bf{r}$ is the received discrete signal vector, and ${\bf{b}}_{{\rm{i}},{\rm{MF}}}^{\rm{D}}$ is the estimated DS-CDMA ith symbol vector. 

\subsection{Rake Detector}
\label{sec:1}
Perhaps, the most popular linear user detection is the Rake detector, which consists of multiple parallel chip-delayed SUD fingers. In this paper, we implement the Rake detector with the estimated known channel gain coefficients, but not the channel delays. One can express the Rake detector for the DS-CDMA system mathematically as follows:

\begin{equation}
\bf{b}_{i,Rake}^D =  \bf{S}_i^H{{\bf{H}}^H} \bf{r}
\end{equation}
where $\bf{H}$ represents the estimated channel matrix, and $\bf{b}_{i,Rake}^D$ is the estimated ith user’s symbol vector. 

\subsection{LMMSE Detector}
\label{sec:1}

Conventional linear detectors based on the Least Square (LS), Zero-Force (ZF) and BLUE algorithms [9] perform poorly especially in the presence of colored noise. The LMMSE detector, however, is considered to be one of the best linear detectors for DS-CDMA systems. Mathematically, one can express the LMMSE as follows:

\begin{equation}
\bf{b}_{i,LMMSE}^D =\bf{S}_i^H{{\bf{H}}^H}{\left( {{\sigma ^2}{\bf{H}}{{\bf{H}}^H} + \bf{Q}} \right)^{ - 1}}\bf{r}
\end{equation}
where $\left( {{{\rm{\sigma }}^2}{\bf{H}}{{\bf{H}}^{\rm{H}}} + \bf{Q}} \right) = {\bf{R}} = {\rm{E}}\left[ {{\bf{r}}{{\bf{r}}^{\rm{H}}}} \right]$ is the auto-correlation of the received data at the mobile unit, and ${\sigma ^2}$ is the average power of the received signal. There are several drawbacks in the implementation of the LMMSE receiver. The main drawback is that the computation of the auto-correlation R is very expensive. If possible, one may use eigen-structure decomposition instead of inverting the auto-correlation matrix $\bf{R}$ directly to obtain

\begin{equation}
\bf{b}_{i,LMMSE}^W = \bf{S}_i^H{{\bf{H}}^H}\left( {{\bf{V}_s}\bf{D}_s^{ - 1}\bf{V}_s^H} \right)\bf{r}
\end{equation}
where $\bf{V}_s$ is the estimated eigen-vector matrix of the auto-correlation matrix $\bf{R}$, and $\bf{D}_s$ is the corresponding diagonal eigenvalue matrix. Additionally, one can use adaptive algorithms to estimate the LMMSE user’s symbols as in [32].

\section{THE PROPOSED ADAPTIVE BLIND DETECTION SCHEMES }
\label{sec:1}

In this section, we introduce new blind detection strategies for the filtering structures. We propose three blind multiuser detectors based on (i) a feed-forward structure, (ii) a feedback structure I, and (iii) a feedback structure II, as in [6]. These filtering structures are depicted in figures 3, 4 and 5, respectively.

To that end, one recalls the discrete received signal model (3), namely, 

\[{\bf{ r}_n} = {\bf{ H}_0}{\bf{ b}_n} + {\bf{ H}_1}{\bf{ b}_{n - 1}} + {\bf{ n}_n}\]
The aim here is to detect the symbol vector ${\bf{ b}_n}$  from the received data vector ${\bf{ r}_n}$,  over the discrete index n, under the following assumptions: 

\begin{list}{*}{}
\item   AS1) the G1xK matrices ${\bf{ H}_0}$ and ${\bf{ H}_1}$  are of full column rank.
\item	AS2) the symbol signal vector series,${\bf{ b}_n}$, have statistically independent components and are identically distributed (i.i.d). 
\item	AS3) the Additive Noise vector ${\bf{ n}_n}$ is white, Gaussian, and independent of the symbol source signals.
\item	AS4) the power of the transmitted symbol signals are normalized to be unity.
\item	AS5) the maximum lag in the entire multipath channels is smaller than the spreading gain G of the CDMA  codes.
\item	AS6) the CDMA system is not over-saturated, which means the number of users (K) is less than the number of the spreading gain (G). 
\item	AS7) the channel is assumed to be a slowly fading wide sense stationary.
\end{list}

For methodical convenience, each detector algorithm involves two steps: first, a preprocessing stage; second, the (matrix) rotation stage based on the filtering structures. In the next subsection, we will present the common preprocessing stage (i.e., whitening processes), and then we will derive each of the three algorithms based on each filtering structure in individual subsections. 

\subsection{Step1: Preprocessing (i.e. Data Whitening) }
\label{sec:1}
The outcome of this step is that the symbol signals are detected up to a unitary rotational matrix. This step uses second order statistics (SOS) in order to normalize the variance (or power) of the received discrete signal vector. It may also be used to eliminate redundancy in the data based on PCA. Under Assumptions AS1-AS4, the G1xG1 covariance matrix, say (${\bf{Cov}}$), of the noiseless received discrete signal vector can be expressed as 

\begin{equation}
{\bf{Cov}} = {\bf{E}}\left[ {{{\bf{r}}_n}\bf{r}_n^H} \right] - {q}{{\bf{I}}_{{{G1}}}}
\end{equation}

We will now consider the two-tap signal model. Then we may generalize it using induction techniques. Under Assumptions AS1-AS7, substituting $\bf{r}_n$ from eqn (3) into (13) results in the following covariance matrix:
\begin{equation}
\begin{array}{l}
{\bf{Cov}} = {{\bf{H}}_0}{{E}}\left[ {{{\bf{b}}_n}{\bf{b}}_n^H} \right]{{\bf{H}}_0}^{\bf{H}} + {{\bf{H}}_1}{{E}}\left[ {{{\bf{b}}_{n - 1}}{\bf{b}}_{n - 1}^H} \right]{{\bf{H}}_1}^{\bf{H}}\\
{\bf{Cov}} = \;{{\bf{H}}_0}{{\bf{H}}_0}^{\bf{H}} + {{\bf{H}}_1}{{\bf{H}}_1}^{\bf{H}} = \;{[{{\bf{H}}_0}\;\;\;{{\bf{H}}_1}\left] \; \right[{{\bf{H}}_0}\;\;{{\bf{H}}_1}]^{\bf{H}}}\;
\end{array}
\end{equation}
Observe that under AS2, $E\left[ {{{\bf{b}}_n}{\bf{b}}_n^H} \right] = {{\bf{I}}_{\bf{K}}}$ and $\left[ {{{\bf{b}}_{n - 1}}{\bf{b}}_{n - 1}^H} \right] = {{\bf{I}}_{\bf{K}}}$. Without loss of generality, we shall briefly proceed with the basic algebraic procedure by adopting the eigen-structure decomposition for the symmetric square matrix ${\bf{Cov}}$ and use it to obtain a singular value decomposition for the combined matrix $[{{\bf{H}}_0}\;\;{{\bf{H}}_1}]$ Thus, let

\begin{equation}
{\bf{Cov}} = {\bf{VD}}{{\bf{V}}^{\bf{H}}}
\end{equation}
where ${\bf{V}}$  is a  ${\rm{G}}1{\rm{xG}}1$  matrix of orthogonal eigenvectors satisfying
\begin{equation}
{\bf{V}}{{\bf{V}}^{\bf{H}}} = {{\bf{V}}^{\bf{H}}}{\bf{V}} = {{\bf{I}}_{{\bf{G}}1}}
\end{equation}
and $\bf{D}$ is the corresponding   ${\rm{G}}1{\rm{xG}}1$ diagonal eigen-matrix containing its eigenvalue entries along the diagonal. Thus, from (14), the  ${\rm{G}}{\rm{xG}}$ ${\bf{H}_0}$ and ${\bf{H}_1}$ matrices can be represented respectively as 
\begin{equation}
\begin{array}{l}
{{\bf{H}}_0} = {{\bf{V}}_0}{{\bf{\Lambda }}_0}{{\bf{U}}_0}^{\bf{H}}\\
{{\bf{H}}_1} = {{\bf{V}}_1}{{\bf{\Lambda }}_1}{{\bf{U}}_1}^{\bf{H}}
\end{array}
\end{equation}
where ${\bf{V}_0}$  and ${\bf{V}_1}$ are composed of orderly non-overlapping columns of the  ${\rm{G}}1{\rm{xG}}1$ unitary matrix $\bf{V}$.   $\bf{U}_0$ and $\bf{U}_1$ are constant but unknown ${\rm{K}}{\rm{xK}}$  right singular-value unitary matrices with  ${{\bf{U}}_{{0}}}{{\bf{U}}_{{0}}}^{\bf{H}} = \;{{\bf{U}}_1}{{\bf{U}}_1}^{\bf{H}} = {{\bf{I}}_{\bf{K}}}$, and  ${{\bf{\Lambda }}_0}$ and ${{\bf{\Lambda }}_1}$ are the appropriate ${\rm{G}}{\rm{x K}}$  singular value matrices. We note that the whitening or algebraic PCA procedure can (i) estimate the noise power in eqn(13) and (ii) reduce the whitened signal dimension to the signal subspace, in this case K. Now, we process the received data to obtain the (whitened) data, specifically, we define:

\begin{equation}
{\bf{r}}_n^w = {{\bf{\Lambda }}^ + }{{\bf{V}}^{\bf{H}}}{\bf{r}_n}
\end{equation}
where the ${\rm{K}}{\rm{x G}}1$ matrix ${{\bf{\Lambda }}^ + }$ denotes the pseudo-inverse of the singular value matrices. One simplifies (18) to eventually obtain:	

\begin{equation}
{\bf{r}}_{\bf{n}}^{\bf{w}} = {{\bf{U}}_0}^{\bf{H}}{{\bf{b}}_{\bf{n}}} + {{\bf{U}}_1}^{\bf{H}}{{\bf{b}}_{{\bf{n}} - 1}} + \left( {{{\bf{\Lambda }}^ + }{{\bf{V}}^{\bf{H}}}} \right){{\bf{n}}_{\bf{n}}}
\end{equation}

Thus, the whitening step renders the whitened data expressed in (18) or (19) as having a reduced dimension to the symbol space and a covariance matrix equal to the identity. That is $E\left[ {{\bf{r}}_{\bf{n}}^{\bf{w}}{\bf{r}}_{\bf{n}}^{{\bf{wH}}}} \right] = {\bf{I}_K}$.

Note that, after the preprocessing step, the detection of the symbol signal ${\bf{\hat b}}{{\rm{\;}}_n}$  reduces to determining or compensating for the unknown K x K (rotation) unitary matrices $\bf{U}_0$ and $\bf{U}_1$.  Next, we proceed with the development and derivations of the three proposed adaptive filtering structures, based on (i) feed-forward structure (FF), (ii) feedback structure I (FB-I) and (iii) feedback structure II (FB-II) [6].
\newline

{\bf{Remark:} }For the purposes of the adaptive filtering to be discussed next, we shall re-label these unknown (but fixed) unitary matrices as the starred values for the environment. Specifically, in eqn(19), we set

\[\begin{array}{l}
{\bf{U}_0} = \bf{U}_0^*\\
{\bf{U}_1} = \bf{U}_1^*
\end{array}\]
The developed adaptive filtering will have parameter matrices that, when adaptation is successful, will converge to (approximately) these fixed starred environment parameters. 

\subsection{Step 2a: Determining the rotation unitary matrix $\bf{U}$ for the feedforward structure}
\label{sec:1}

\begin{figure}
 \includegraphics{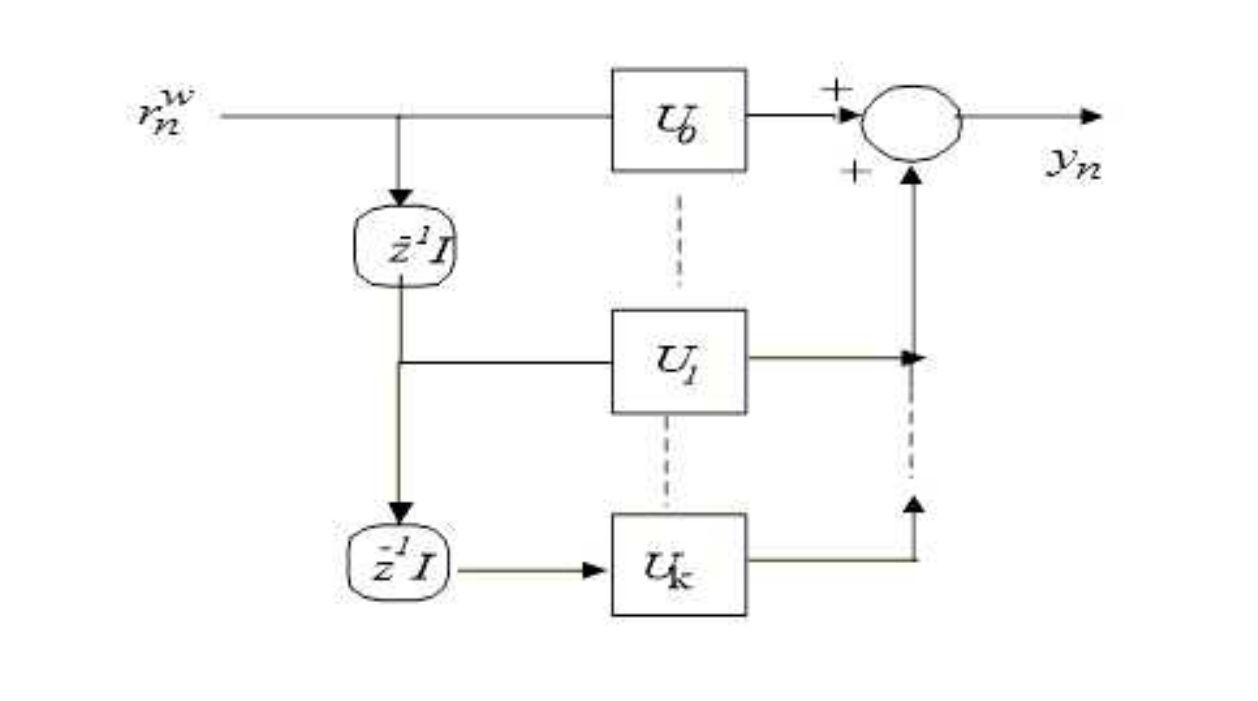}
\caption{Feed-Forward (FF) Demixing Structure}
\label{fig:1}       
\end{figure} 

The output from the FF structure, as depicted in Fig. 3, is expressed as 

\begin{equation}
{\bf{y}_n} = {{\bf{U}}_0} \bf{r}_n^w + \mathop \sum \limits_{k = 1}^K {{\bf{U}}_k} \bf{r}_{n - k}^w
\end{equation}
For simplicity of presentation, we begin with a “two-tap” model; thus the two-tap FF structure becomes 

\begin{equation}
{\bf{y}_n} = {{\bf{U}}_0}r_n^w + {{\bf{U}}_1}\bf{r}_{n - 1}^w
\end{equation}

The goal for a successful adaptive algorithm is to bring about the convergence of the parameter matrices to the (starred) environment parameters. Specially, the adaptive algorithm succeeds when its parameter matrices converge to
$\bf{U}_0^*$, and $\bf{U}_1^*$  respectively.

We now proceed with the development. One can re-write this convolutive filter (21) as the following (static) map
\begin{equation}
\left[ {\begin{array}{*{20}{c}}
{{\bf{y}_n}}\\
{\bf{r}_{n - 1}^w}
\end{array}} \right] = \left[ {\begin{array}{*{20}{c}}
{{\bf{U}_0}\;\;\;\;\;\;\;\;{\bf{U}_1}}\\
{\bf{0}\;\;\;\;\;\;\;\;\;\;\;\;\;\bf{I}}
\end{array}} \right]\left[ {\begin{array}{*{20}{c}}
{\bf{r}_n^w}\\
{\bf{r}_{n - 1}^w}
\end{array}} \right]
\end{equation}
Then, one defines, respectively, the new augmented output, the static map, and the augmented input as

\[
{\rm{\;}}\tilde{\bf{Y}} = \left[ {\begin{array}{*{20}{c}}
{{\bf{y}_n}}\\
{\bf{r}_{n - 1}^w}
\end{array}} \right] \]

\[
\tilde{\bf{U}} = \left[ {\begin{array}{*{20}{c}}
{{\bf{U}_0}\;\;\;\;\;\;\;\;\;\bf{0}}\\
{\;{\bf{U}_1}\;\;\;\;\;\;\;\;\;\;\bf{I}}
\end{array}} \right] \]

\[
\tilde{\bf{R}} = \left[ {\begin{array}{*{20}{c}}
{\bf{r}_n^w}\\
{\bf{r}_{n - 1}^w}
\end{array}} \right] \]
Thus, the expression in (22) becomes the static map

\begin{equation}
\tilde{\bf{Y}} = {\tilde{\bf{U}}^H}\tilde{\bf{R}}
\end{equation}

Based on the natural gradient approach [43, 44], the update law for the  columns of the augmented de-mixing matrix $\tilde{\bf{U}}$ can be expressed as

\begin{equation}
{\bf{u}^ + } = \bf{u} - \mu E\left[ {\tilde {\bf{R}}\left( {g\left( {{\bf{u}^H}\tilde{ \bf{R}}}\right)} \right)} \right]
\end{equation}
where ${\bf{u} }$, respectively ${\bf{u}^ + }$, is the current, respectively next, value of one column vector of  ${\tilde{\bf{U}}}$, $\mu$ is the step size and g is the chosen score function. Noting the structure of the de-mixing matrix in (23), one decomposes the column vector as 
\begin{equation}
\bf{u} = \left[ {\begin{array}{*{20}{c}}
{{\bf{u}_0}}\\
{{\bf{u} _1}}
\end{array}} \right]
\end{equation}
Hence, the update law is correspondingly decomposed as (Note, we have suppressed the $E[.]$ operator):
\begin{equation}
\left[ {\begin{array}{*{20}{c}}
{{\bf{u}_0}^ + }\\
{{\bf{u}_1}^ + }
\end{array}} \right] = \left[ {\begin{array}{*{20}{c}}
{{\bf{u}_0}}\\
{{\bf{u}_1}}
\end{array}} \right] - \mu \;\left[ {\begin{array}{*{20}{c}}
{\bf{r}_n^w}\\
{\bf{r}_{n - 1}^w}
\end{array}} \right]g\left( {{\bf{y}_n}} \right)
\end{equation}
where $\bf{u}_0$ , $\bf{u}_1$ are the column vectors of $\bf{U}_0$ and $\bf{U}_1$ in (22), respectively. Therefore, the update laws for the individual (sub-) columns are 
\begin{equation}
{\bf{u}_0}^ +  = {\bf{u}_0} - \mu \bf{r}_n^wg\left( {{\bf{y}_n}} \right)
\end{equation}

\begin{equation}
{\bf{u}_1}^ +  = {\bf{u}_1} - \mu \bf{r}_{n-1}^wg\left( {{\bf{y}_n}} \right)
\end{equation}
Now, by induction, the update law for the kth lag element $\bf{u}_k$ is 

\begin{equation}
{\bf{u}_k}^ +  = {\bf{u}_k} - \mu \bf{r}_{n-k}^wg\left( {{\bf{y}_n}} \right)
\end{equation}

\subsection{Step 2b: Determining the rotation unitary matrix $\bf{U}$ based on feedback structure I (FB-I)}
\label{sec:1}

\begin{figure}
 \includegraphics{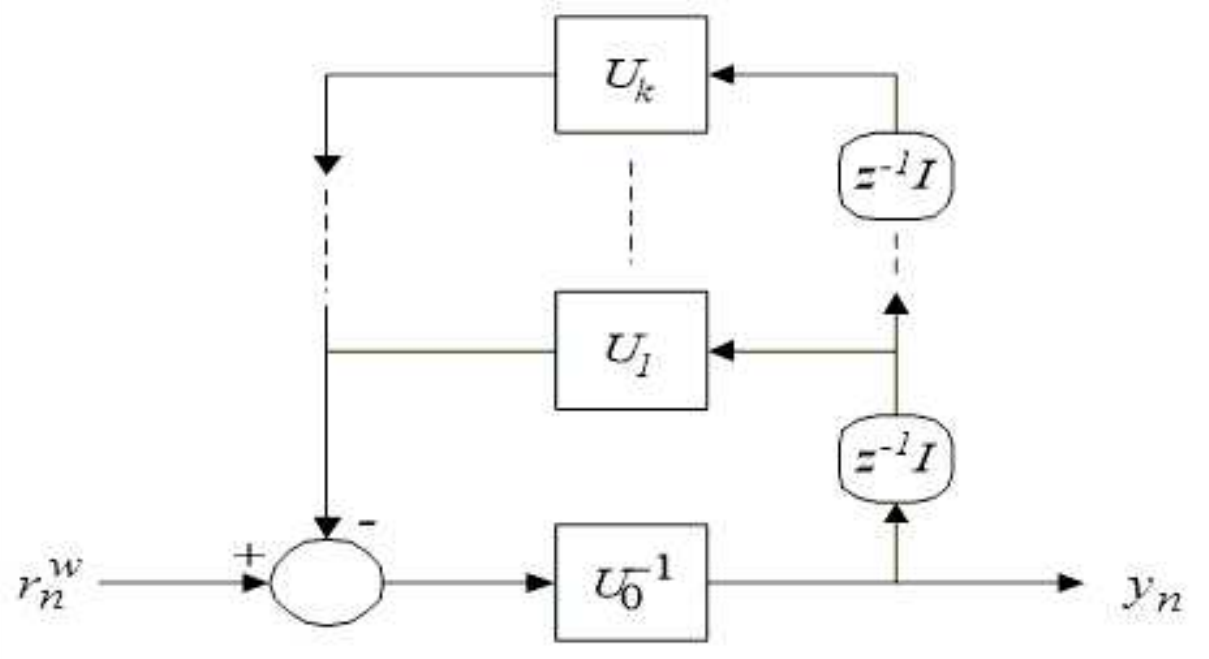}
\caption{Feedback Demixing Structure I FB-I}
\label{fig:1}       
\end{figure} 

The output of  FB-I, as depicted in Fig. 4, results in the filtering expression  
\begin{equation}
{\bf{y}_n} = \bf{U}_0^{ - 1}\left( {\bf{r}_n^w - \mathop \sum \limits_{k = 1}^K {\bf{U}_k}{\bf{y}_{n - k}}} \right)
\end{equation}
Consider now just two taps of FB-I , i.e.,  

\begin{equation}
{\bf{y}_n} = {\bf{U}_0}^{ - 1}\left( {\bf{r}_n^w - {\bf{U}_1}{\bf{y}_{n - 1}}} \right)
\end{equation}
One can re-write this convolutive filter into the following augmented static form
\begin{equation}
\left[ {\begin{array}{*{20}{c}}
{\bf{r}_n^w}\\
{{\bf{y}_{n - 1}}}
\end{array}} \right] = \left[ {\begin{array}{*{20}{c}}
{{\bf{U}_0}\;\;\;\;\;\;\;\;{\bf{U}_1}}\\
{\bf{0}\;\;\;\;\;\;\;\;\;\;\;\;\;\bf{I}}
\end{array}} \right]\left[ {\begin{array}{*{20}{c}}
{{\bf{y}_n}}\\
{{\bf{y}_{n - 1}}}
\end{array}} \right]
\end{equation}
Or

\[\left[ {\begin{array}{*{20}{c}}
{{\bf{y}_n}}\\
{{\bf{y}_{n - 1}}}
\end{array}} \right] = {\left[ {\begin{array}{*{20}{c}}
{{\bf{U}_0}\;\;\;\;\;\;\;\;{\bf{U}_1}}\\
{\bf{0}\;\;\;\;\;\;\;\;\;\;\;\;\;\bf{I}}
\end{array}} \right]^{ - 1}}\left[ {\begin{array}{*{20}{c}}
{\bf{r}_n^w}\\
{{\bf{y}_{n - 1}}}
\end{array}} \right]\]

\begin{equation}
\left[ \begin{array}{l}
{\bf{y}_n}\\
{\bf{y}_{n - 1}}
\end{array} \right] = \left[ \begin{array}{l}
\bf{U}_0^{ - 1}\\
  \;\;\;\bf{0}
\end{array} \right.\left. \begin{array}{l}
 - \bf{U}_0^{ - 1}{\bf{U}_1}\\
     \;\;\;\;\;\; \bf{I}
\end{array} \right]\left[ \begin{array}{l}
\bf{r}_n^w\\
{\bf{y}_{n - 1}}
\end{array} \right]
\end{equation}
Thus, in this case, one defines the augmented “output, de-mixing matrix, and input” as follows: 

\[
\tilde{\bf{Y}} = \left[ {\begin{array}{*{20}{c}}
{{\bf{y}_n}}\\
{{\bf{y}_{n - 1}}}
\end{array}} \right]
\]

\[
\tilde{\bf{U}} = \left[ {\begin{array}{*{20}{c}}
{{\bf{U}_0}\;\;\;\;\;\;\;\;\;\;\;\bf{0}}\\
{{\bf{U}_1}\;\;\;\;\;\;\;\;\;\;\;\bf{I}}
\end{array}} \right]
\]

\[
\tilde{\bf{R}} = \left[ {\begin{array}{*{20}{c}}
{\bf{r}_n^w}\\
{{\bf{y}_{n - 1}}}
\end{array}} \right]
\]
One then re-expresses (32) into the compact equation
\begin{equation}
\tilde{\bf{R}} = {\tilde{\bf{U}}^H}\tilde{\bf{Y}}
\end{equation}
Again, using the natural gradient approach, the update law for a column of the de-mixing matrix $\tilde \bf{U}$ is 

\begin{equation}
{\bf{u}^ + } = \bf{u} - \mu E\left[ {\tilde{\bf{Y}}\left( {g\left( {{\bf{u}^H}\tilde{\bf{Y}}} \right)} \right)} \right]
\end{equation}
As before, ${\bf{u} }$, respectively ${\bf{u}^ + }$, is the current, respectively next, value of one column vectors of  ${\tilde{\bf{U}}}$, $\mu$ is the step size and g is the chosen score function. 
\newline

One can exploit the block matrix structure of the de-mixing matrix and simplify the update law. To that end, consider the block matrix

\begin{equation}
{\bf{u}^0} = \left[ {\begin{array}{*{20}{c}}
{{{\bf{u}_0}^0}}\\
{ {\bf{u}_1}^0}
\end{array}} \right]
\end{equation}
Thus, the update laws can be calculated to produce

\begin{equation}
\left[ {\begin{array}{*{20}{c}}
{{{\bf{u}_0}^ + }}\\
{{\bf{u}_1}^ + }
\end{array}} \right] = \left[ {\begin{array}{*{20}{c}}
{{\bf{u}_0}}\\
{{\bf{u}_1}}
\end{array}} \right] - \;\mu \left[ {\begin{array}{*{20}{c}}
{\bf{y}_n}\\
{{\bf{y}_{n - 1}}}
\end{array}} \right]g\left( {{\bf{u}_0}\bf{y}_n + {\bf{u}_1}{\bf{y}_{n - 1}}} \right)
\end{equation}
Similarly, the next block matrices can be defined as
\begin{equation}
{\bf{u}^1} = \left[ {\begin{array}{*{20}{c}}
{{\bf{0}}}\\
{{\bf{i}^1}}
\end{array}} \right]
\end{equation}
This leads to the specialized form 

\begin{equation}
\left[ {\begin{array}{*{20}{c}}
{{\bf{0}^ + }}\\
{{\bf{i}^ + }}
\end{array}} \right] = \left[ {\begin{array}{*{20}{c}}
\bf{0}\\
{{\bf{i}}}
\end{array}} \right] - \mu \;\left[ {\begin{array}{*{20}{c}}
{\bf{y}_n}\\
{{\bf{y}_{n - 1}}}
\end{array}} \right]g\left( {{\bf{y}_{n - 1}}} \right)
\end{equation}
Thus, the update laws for the individual columns are  
\begin{equation}
{\bf{u}_0}^ +  = {\bf{u}_0} - \mu {\bf{y}_{n}}g\left( {\bf{r}_n^w} \right)
\end{equation}
and
\begin{equation}
{\;\;\;\;\;\; \bf{u}_1}^ +  = {\bf{u}_1} - \mu {\bf{y}_{n - 1}}g\left( {\bf{r}_n^w}  \right)
\end{equation}
Analogously, by induction, the update law for the kth lag element, say $\bf{u}_k$  is 

\begin{equation}
{\;\;\;\;\;\;\;\;\; \bf{u}_k}^ +  = {\bf{u}_k} - \mu {\bf{y}_{n - k}}g\left( {\bf{r}_n^w} \right)
\end{equation}

\subsection{Step 2c: Determining the rotation unitary matrix  $\bf{U}$ based on Feedback Structure II}
\label{sec:1}

\begin{figure}
 \includegraphics{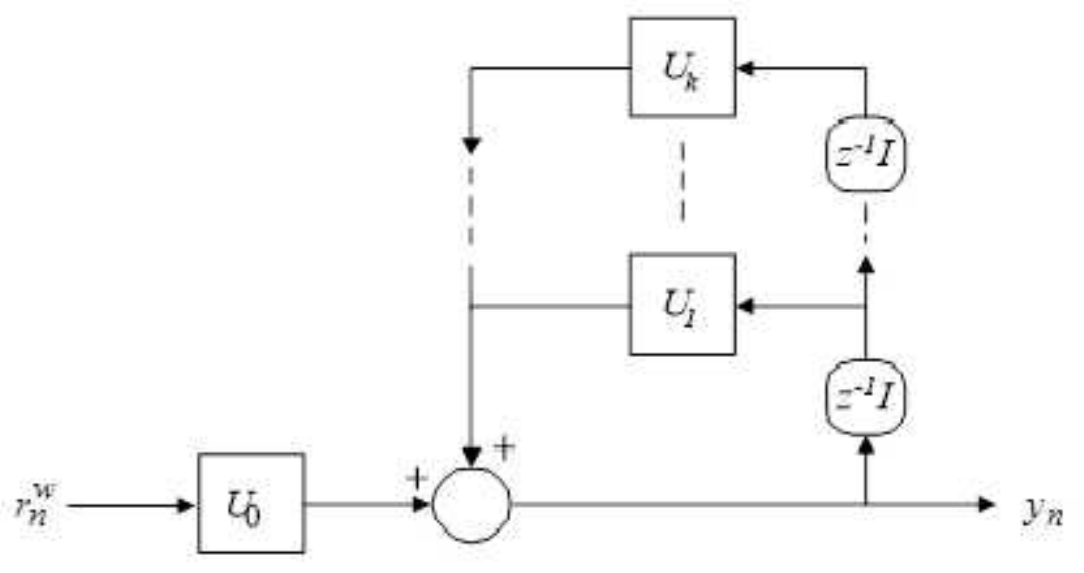}
\caption{Feedback Demixing Structure II (FB-II)}
\label{fig:1}       
\end{figure} 

The output of FB-II, as depicted in Fig. 5, is expressed as:
\begin{equation}
{\bf{y}_n} = {\bf{U}_0}r_n^w + \mathop \sum \limits_{k = 1}^K {\bf{U}_k}{\bf{y}_{n - k}}
\end{equation}
Again, consider two taps of FB-II, i.e.,

\begin{equation}
{\bf{y}_n} = {\bf{U}_0}\bf{r}_n^w - {\bf{U}_1}{\bf{y}_{n - 1}}
\end{equation}
Hence, one re-writes this convolutive filter in the following augmented static form
\begin{equation}
\left[ {\begin{array}{*{20}{c}}
{{\bf{y}_n}}\\
{{\bf{y}_{n - 1}}}
\end{array}} \right] = \left[ {\begin{array}{*{20}{c}}
{{\bf{U}_0}\;\;\;\;\;\;\;\; - {\bf{U}_1}}\\
{\bf{0}\;\;\;\;\;\;\;\;\;\;\;\;\;\bf{I}}
\end{array}} \right]\left[ {\begin{array}{*{20}{c}}
{\bf{r}_n^w}\\
{{\bf{y}_{n - 1}}}
\end{array}} \right]
\end{equation}
Similarly, define the augmented entities as 

\[
\tilde{\bf{Y}} = \left[ {\begin{array}{*{20}{c}}
{{\bf{y}_n}}\\
{{\bf{y}_{n - 1}}}
\end{array}} \right]
\]

\[
\tilde{\bf{U}} = \left[ {\begin{array}{*{20}{c}}
{\;{\bf{U}_0}\;\;\;\;\;\;\;\;\bf{0}}\\
{ - {\bf{U}_1}\;\;\;\;\;\;\;\bf{I}}
\end{array}} \right]
\]

\[
\tilde{\bf{R}} = \left[ {\begin{array}{*{20}{c}}
{\bf{r}_n^w}\\
{{\bf{y}_{n - 1}}}
\end{array}} \right]
\]
Thus one re-writes (45) into the compact mapping  

\begin{equation}
\tilde{\bf{Y}} = {\tilde{\bf{U}}^H}\tilde{\bf{R}}
\end{equation}

Using the natural gradient approach, the update laws for a weight column of the de-mixing matrix $\tilde{\bf{U}}$ is expressed as

\begin{equation}
{\bf{u}^ + } = \bf{u} - \mu E\left[ {\tilde{\bf{R}}\left( {g\left( {{\bf{u}^H}\tilde{\bf{R}} }\right)} \right)} \right]
\end{equation}
where, as before, ${\bf{u} }$, respectively ${\bf{u}^ + }$, is the current, respectively next, value of one column vectors of  ${\tilde{\bf{U}}}$, $\mu$ is the step size and $\bf{g(.)}$ is the chosen score function. One can appropriately decompose a column vector in order to simplify the update expressions as: 

\begin{equation}
\bf{u} = \left[ {\begin{array}{*{20}{c}}
{{\bf{u}_0}}\\
{{\bf{u}_1}}
\end{array}} \right]
\end{equation}
Then the update law becomes decomposed as follows:

\begin{equation}
\left[ {\begin{array}{*{20}{c}}
{{\bf{u}_0}^ + }\\
{{\bf{u}_1}^ + }
\end{array}} \right] = \left[ {\begin{array}{*{20}{c}}
{{\bf{u}_0}}\\
{{\bf{u}_1}}
\end{array}} \right] - \mu \;\left[ {\begin{array}{*{20}{c}}
{\bf{r}_n^w}\\
{{\bf{y}_{n - 1}}}
\end{array}} \right]g\left( {{\bf{y}_n}} \right)
\end{equation}
Thus, the update laws for the individual sub-columns are 

\begin{equation}
{\bf{u}_0}^ +  = {\bf{u}_0} - \mu \bf{r}_n^wg\left( {{\bf{y}_n}} \right)
\end{equation}
and 

\begin{equation}
{\bf{u}_1}^ +  = {\bf{u}_1} - \mu {\bf{y}_{n - 1}}g\left( {{\bf{y}_n}} \right)
\end{equation}
Finally, by induction, the update law for the kth lag element $\bf{u}_k$ is

\begin{equation}
{\bf{u}_k}^ +  = {\bf{u}_k} - \mu {\bf{y}_{n - k}}g\left( {{\bf{y}_n}} \right)
\end{equation}

\subsection{The proposed adaptive Rake-based detectors}
\label{sec:1}

While the previous filtering structures constitute new adaptive filters, one can augment the existing conventional Rake detectors to improve its performance adaptively.   We now develop three adaptive modifications of the conventional Rake detector based on, respectively, Independent Component Analysis (ICA), Robust ICA and Principle Component Analysis (PCA). Recalling the Rake detector's structure as given in (10), one can mathematically express the adaptive modified Rake detector for DS-CDMA systems as follows:
\begin{equation}
\bf{b}_{i,Rake}^D = \bf{S}_i^H{\bf{W}}{\bf{H}^H}\bf{r}
\end{equation}
where, as before, $\bf{H}$ is the crudely estimated (inverse) channel matrix usually based only on time-delays of the “finger” Rake filter, $\bf{S}_i$ is a vector associated with the ith user's signature code, and  $\bf{b}_{i,Rake}^D$ is the estimated ith user's symbol. A $ GxG$ matrix $\bf{W}$ is inserted which will adaptively augment and improve the estimate of the channel inverse.  In the following, we summarize the process in $\bf{Algorithms}$ 1, 2, and 3 to adaptively estimate the matrix $\bf{W}$  using the FastICA, Robust ICA and PCA algorithms, respectively.

\makeatletter
\def\BState{\State\hskip-\ALG@thistlm}
\makeatother

\begin{algorithm}
\caption{Adaptive Rake based FastICA method}\label{euclid}
\begin{algorithmic}[1]
\Procedure{ Initialization}{}
\State $\textit{$\bf{r}$} \gets \text{$M\; \times \;N$ matrix of realization }$
\State $\textit{$\;\bf{W} = {\bf{I}_G}$} \gets \text{Initial demixing matrix }$
\State $\textit{Itr} \gets \text{number of  }\textit{iterations}$
\State $\textit{$\gamma $} \gets \text{Step Size }$
\State $\textit{$\bf{H}$} \gets \text{the estimated channel matrix}$
\State $\textit{$g\left( \bf{y} \right) = {\bf{y}^3}$} \gets \text{the nonlinear fcn}$
\BState \emph{Pre-Whitening}:
\State $\textit{$\bf{r}$} \gets \text{${\;\bf{V}*\bf{r} = {\Lambda ^{\left( {\left( { - 1} \right) / 2} \right)}}\;{\bf{E}^T}\;\;\bf{r}}$ }$
\BState \emph{For Loop}:
\State $\textit{i} \gets \text{$1 \ldots N$ }$
\State $\textit{$\bf{r}$} \gets \text{$\bf{W}{\bf{H}^H}\bf{r}\left( {:,i} \right)$}$
\State $\textit{${\bf{W}^ + }$} \gets \text{$E[{\left[ {g\left( {\bf{Wr}} \right)} \right]^T}] - E[g'\left( {\bf{Wr}} \right)]\bf{W}\;$}$
\BState \emph{Normalization}:
\State $\textit{${\bf{W}^ + }$} \gets \text{$\bf{W}/norm\left( \bf{W} \right)$}$
\State $\textit{$\bf{b}_{i,ICA}^D\left( {:,i} \right)$} \gets \text{$\bf{S}_i^H{\bf{W}}{\bar \bf{H}^H}\bf{r}$}$
\State \textbf{goto} \emph{For Loop}.
\State \textbf{close};
\BState \emph{Output}:
\State $\textit{$\bf{b}_{i,ICA}^D$} \gets \text{the estimated Symbols }$
\EndProcedure
\end{algorithmic}
\end{algorithm}

\begin{algorithm}
\caption{Adaptive Rake based RICA method}\label{euclid}
\begin{algorithmic}[1]
\Procedure{ Initialization}{}
\State $\textit{$\bf{r}$} \gets \text{$M\; \times \;N$ matrix of realization }$
\State $\textit{$\;\bf{W} = {\bf{I}_G}$} \gets \text{Initial demixing matrix }$
\State $\textit{Itr} \gets \text{number of  }\textit{iterations}$
\State $\textit{$\mu $} \gets \text{Step Size }$
\State $\textit{$\bf{H}$} \gets \text{the estimated channel matrix}$
\State $\textit{$g\left( \bf{y} \right) $} \gets \text{the gradient of the Kurtosis}$
\BState \emph{Pre-Whitening}:
\State $\textit{$\bf{r}$} \gets \text{${\;\bf{V}*\bf{r} = {\Lambda ^{\left( {\left( { - 1} \right) / 2} \right)}}\;{\bf{E}^T}\;\;\bf{r}}$ }$
\BState \emph{For Loop}:
\State $\textit{i} \gets \text{$1 \ldots N$ }$
\State $\textit{$\bf{r}$} \gets \text{$\bf{W}{\bf{H}^H}\bf{r}\left( {:,i} \right)$}$
\State $\textit{${\bf{W}^ + }$} \gets \text{$\bf{W} + \;\mu \left( {{\bf{I}_G} - g\left(\bf{r} \right)*g{{\left(\bf{ r} \right)}^H}} \right)\bf{W}$}$
\BState \emph{Normalization}:
\State $\textit{${\bf{W}^ + }$} \gets \text{$\bf{W}/norm\left( \bf{W} \right)$}$
\State $\textit{$\bf{b}_{i,RICA}^D\left( {:,i} \right)$} \gets \text{$\bf{S}_i^H{\bf{W}}{\bar \bf{H}^H}\bf{r}$}$
\State \textbf{goto} \emph{For Loop}.
\State \textbf{close};
\BState \emph{Output}:
\State $\textit{$\bf{b}_{i,RICA}^D$} \gets \text{the estimated Symbols }$
\EndProcedure
\end{algorithmic}
\end{algorithm}

\begin{algorithm}
\caption{Adaptive Rake based PCA method}\label{euclid}
\begin{algorithmic}[1]
\Procedure{ Initialization}{}
\State $\textit{$\bf{r}$} \gets \text{$M\; \times \;N$ matrix of realization }$
\State $\textit{$\;\bf{W} = {\bf{I}_G}$} \gets \text{Initial demixing matrix }$
\State $\textit{Itr} \gets \text{number of  }\textit{iterations}$
\State $\textit{$\gamma $} \gets \text{Step Size }$
\State $\textit{$\bf{H}$} \gets \text{the estimated channel matrix}$
\BState \emph{Pre-Whitening}:
\State $\textit{$\bf{r}$} \gets \text{${\;\bf{V}*\bf{r} = {\Lambda ^{\left( {\left( { - 1} \right) / 2} \right)}}\;{\bf{E}^T}\;\;\bf{r}}$ }$
\BState \emph{For Loop}:
\State $\textit{i} \gets \text{$1 \ldots N$ }$
\State $\textit{$\bf{r}$} \gets \text{$\bf{W}{\bf{H}^H}\bf{r}\left( {:,i} \right)$}$
\State $\textit{${\bf{W}^ + }$} \gets \text{$\bf{W} + \;\gamma \left( {{\bf{I}_G} - \bf{r}*{\bf{r}^H}} \right)\bf{W}$}$
\BState \emph{Normalization}:
\State $\textit{${\bf{W}^ + }$} \gets \text{$\bf{W}/norm\left( \bf{W} \right)$}$
\State $\textit{$\bf{b}_{i,PCA}^D\left( {:,i} \right)$} \gets \text{$\bf{S}_i^H{\bf{W}}{\bar \bf{H}^H}\bf{r}$}$
\State \textbf{goto} \emph{For Loop}.
\State \textbf{close};
\BState \emph{Output}:
\State $\textit{$\bf{b}_{i,PCA}^D$} \gets \text{the estimated Symbols }$
\EndProcedure
\end{algorithmic}
\end{algorithm}

\section{SIMULATION RESULTS}
\label{sec:1}

A series of extensive simulations are carried out in order to verify and evaluate the performance of the proposed adaptive filters and algorithms in the multipath downlink DS-CDMA system in the presence of AWGN. We summarize the case study results as follows. We assume a constant spreading gain, which is $G=63$ for Gold Codes and  $G=64$ for Orthogonal Variable Spreading Factor (OVSF) codes. The received CDMA signal experiences five multipath channels L=5 with delays of {0,1,2,3 ,4} chips, respectively. Also, we set the complex attenuation coefficients to represent the multipath channels, specifically,  $ h_0=0.3684 + 0.5364i$, $ h_1=0.1982 + 0.0187i$, $h_2=0.0237 + 0.5683$, $ h_3=0.1112 + 0.0835i$, and $h_4=0.2203 + 0.2756i $, respectively. We use the following model function for sub-Gaussian sources for which the source signals have a negative kurtosis sign:

\begin{equation}
{{\rm{g}}^{{\rm{SUB}}}}\left( {{\bf{\hat b}}} \right) = {\bf{\hat b}} - \left( {\tanh \left( {{\rm{Re}}\left\{ {{\bf{\hat b}}} \right\}} \right) + {\rm{jtanh}}\left( {{\rm{Im}}\left\{ {{\bf{\hat b}}} \right\}} \right)} \right)
\end{equation}

Monte Carlo Simulations have been run to verify the validity of the algorithms. We also use the signal-to-noise ratio (SNR) as a figure of merit which represents the ratio of the energy per symbol and the power spectral density (PSD) of the noise. Moreover, all the user symbols are assumed to be transmitted with the same power. Fig. 6 (a) and (b) show the simulation results of BER vs. SNR  for the proposed detectors in contrast to  the existing and conventional ones for the number of users K=30 and K=50, respectively. The other parameters were set as (i) number of symbols M=1000 and (ii) number of paths L=5, with the values of SNR in the range of -10 dB to 30dB. 

\begin{figure*}[ht]
\centering
    \begin {subfigure} [{Using 30 users}] {
          \includegraphics[height=200pt,width=300pt,angle=0]{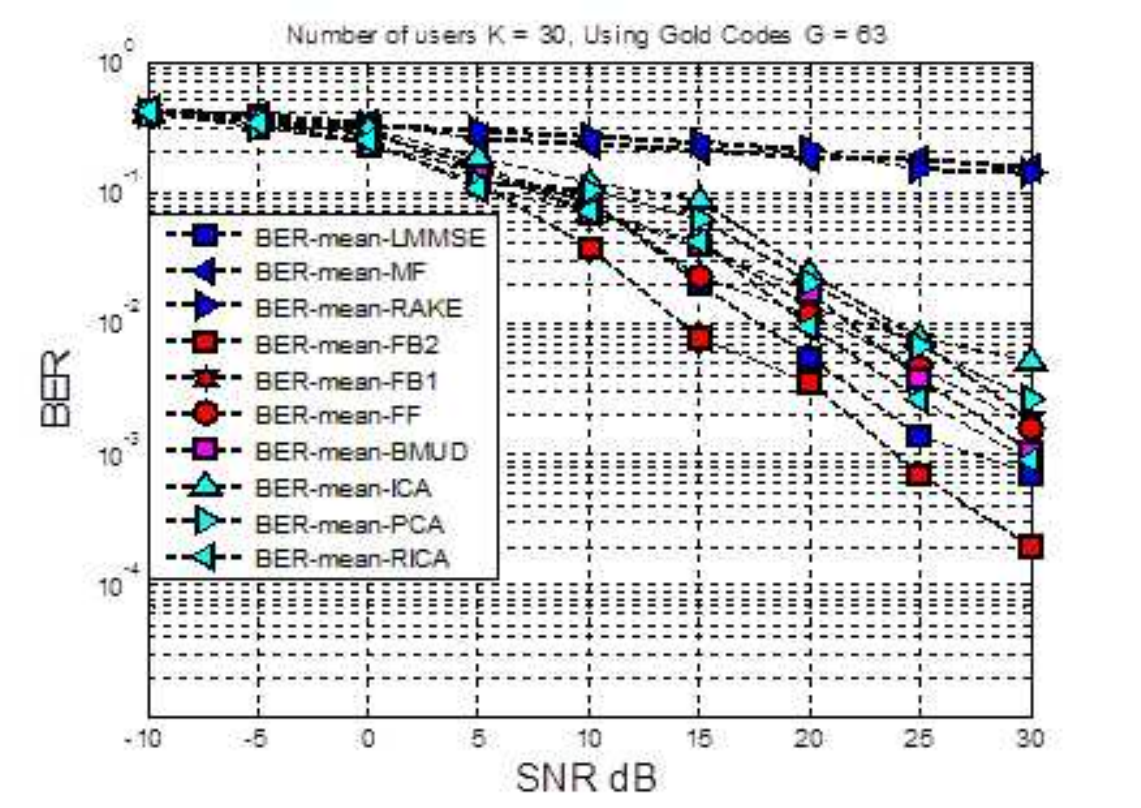}
}
\end {subfigure}

  \begin {subfigure} [{Using 50 users }]{
          \includegraphics[height=200pt,width=300pt,angle=0]{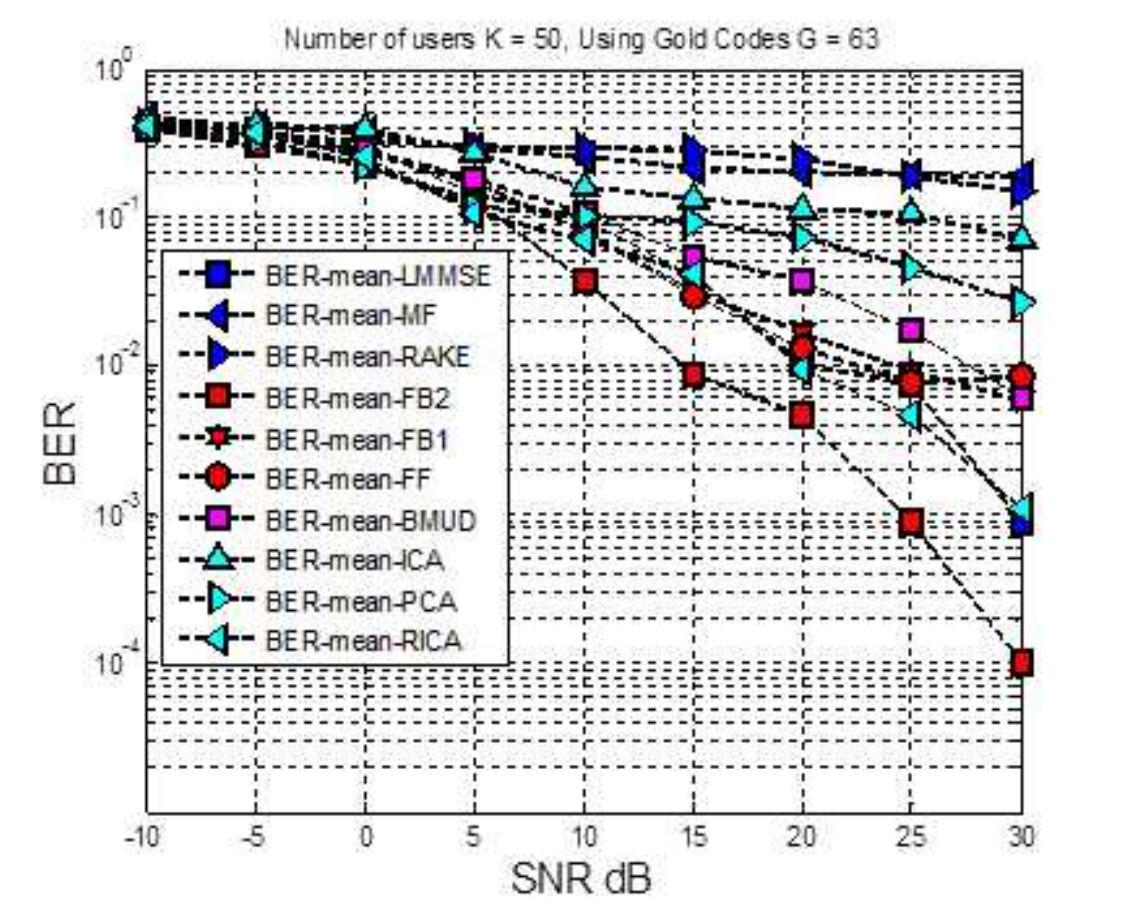}
}
\end {subfigure}

\caption{Average BER as a function of SNR for DS-CDMA downlink. Using Gold codes G=63. (a) Using 30 users (b) Using 50 users}
\label{fig:2}
\end{figure*}

Fig. 6 shows that the proposed algorithms improve the performance of the CDMA system. One observes that the blind multiuser detection based on FB-II has resulted in the lowest BER, and thus it outperforms all other detectors. One also observes that the proposed algorithms work even in cases which cause difficulties for the   LMMSE receiver, as in the high SNR ratio, and when the sample set is fairly small. Moreover, the performance of the blind multiuser detection degrades as the number of users increases as comparatively seen in Fig. 6 (b). 

Furthermore, we have also evaluated the effect of the OVSF codes as depicted in Fig. 7. As in Fig. 7, it is generally the case that using the OVSF codes enhances the performance of the proposed methods. 

\begin{figure*}[ht]
\centering
    \begin {subfigure} [{Using 30 users}] {
          \includegraphics[height=200pt,width=300pt,angle=0]{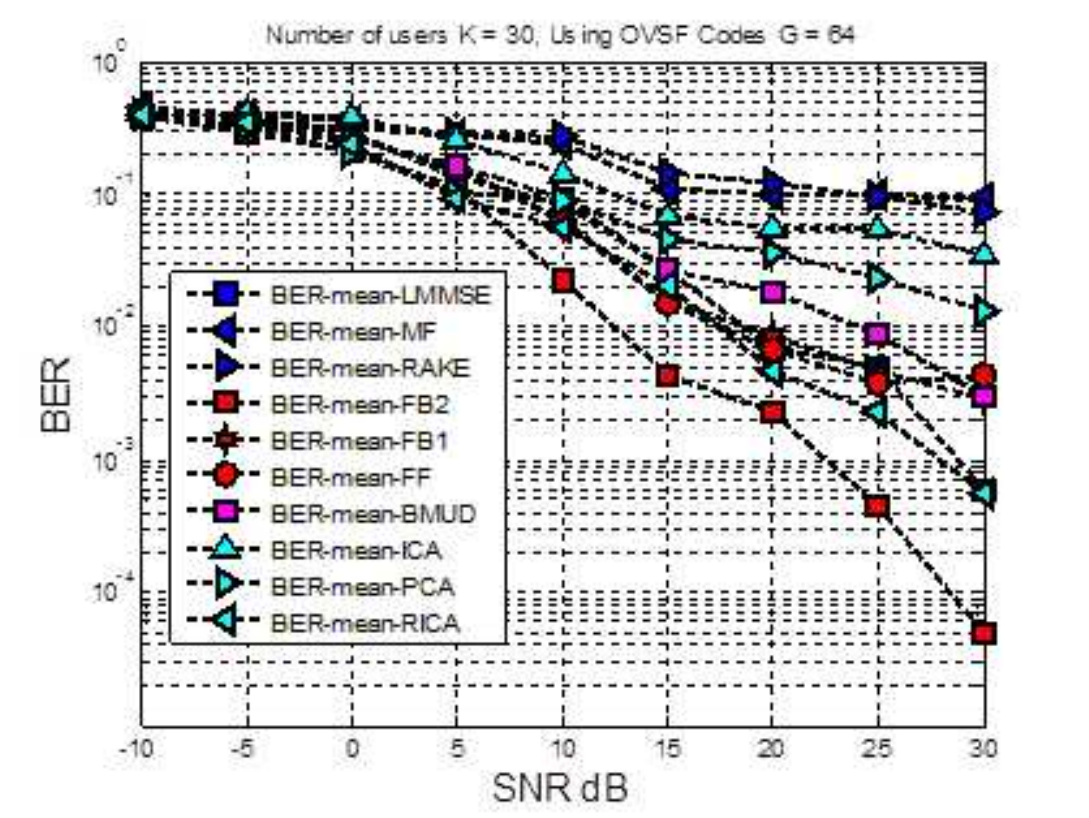}
}
\end {subfigure}

  \begin {subfigure} [{Using 50 users }]{
          \includegraphics[height=200pt,width=300pt,angle=0]{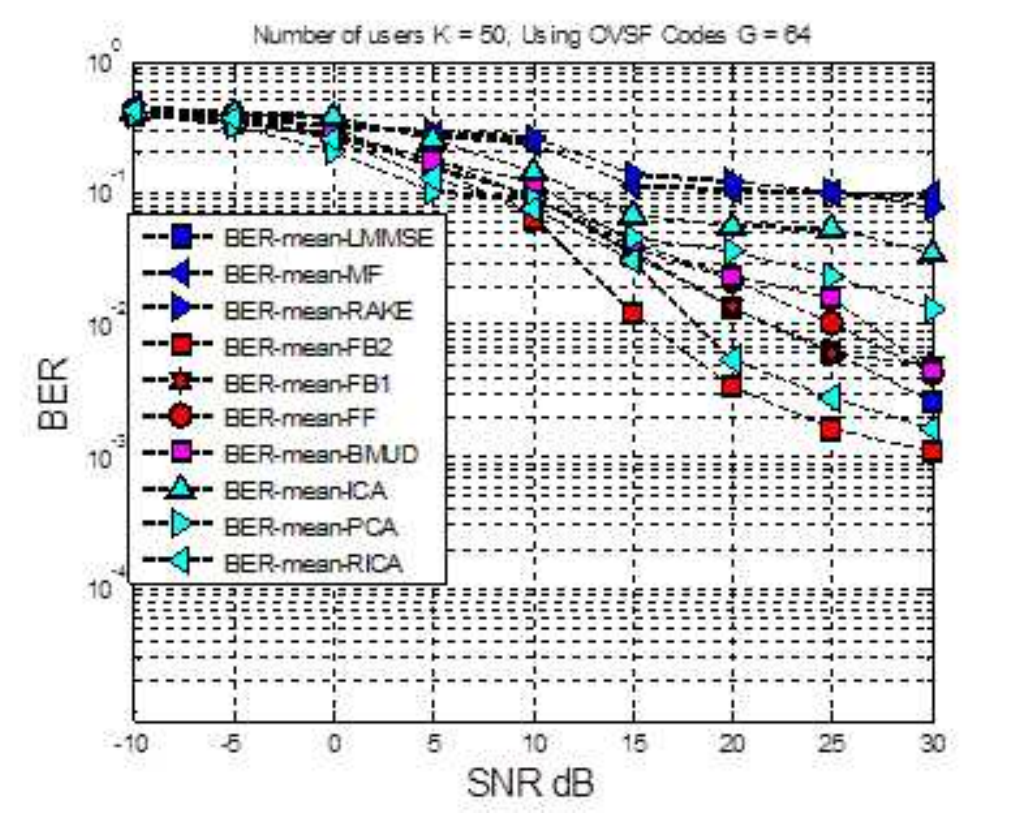}
}
\end {subfigure}
\caption{Average BER as a function of SNR for DS-CDMA downlink. Using OVSF codes G=64. (a) Using 30 users (b) Using 50 users}
\label{fig:2}
\end{figure*}

In the WCDMA System case, we assume that the channel coefficients are $ h_0=0.3684 + 0.5364i$, $ h_1=0.1982 + 0.0187i$, $h_2=0.0237 + 0.5683$, $ h_3=0.1112 + 0.0835i$, and $h_4=0.2203 + 0.2756i $, respectively. Also,  all user-specific codes use two types of spreading codes, namely, Gold codes with spreading gain G=63 and OVSF (or Walsh-Hadamard) codes with spreading gain G=64. 

In Fig. 8 and 9, we document and demonstrate the performance of the various methods in terms of BER for the WCDMA downlink scenario.  We observe that the LMMSE is slightly better than some presented detectors under good SNR conditions. However, the proposed algorithm based on FB-II outperforms all detectors over all SNR depicted ranges and has again produced the lowest BER when compared to all other methods. 

\begin{figure*}[ht]
\centering
    \begin {subfigure} [{Using 30 users}] {
          \includegraphics[height=200pt,width=300pt,angle=0]{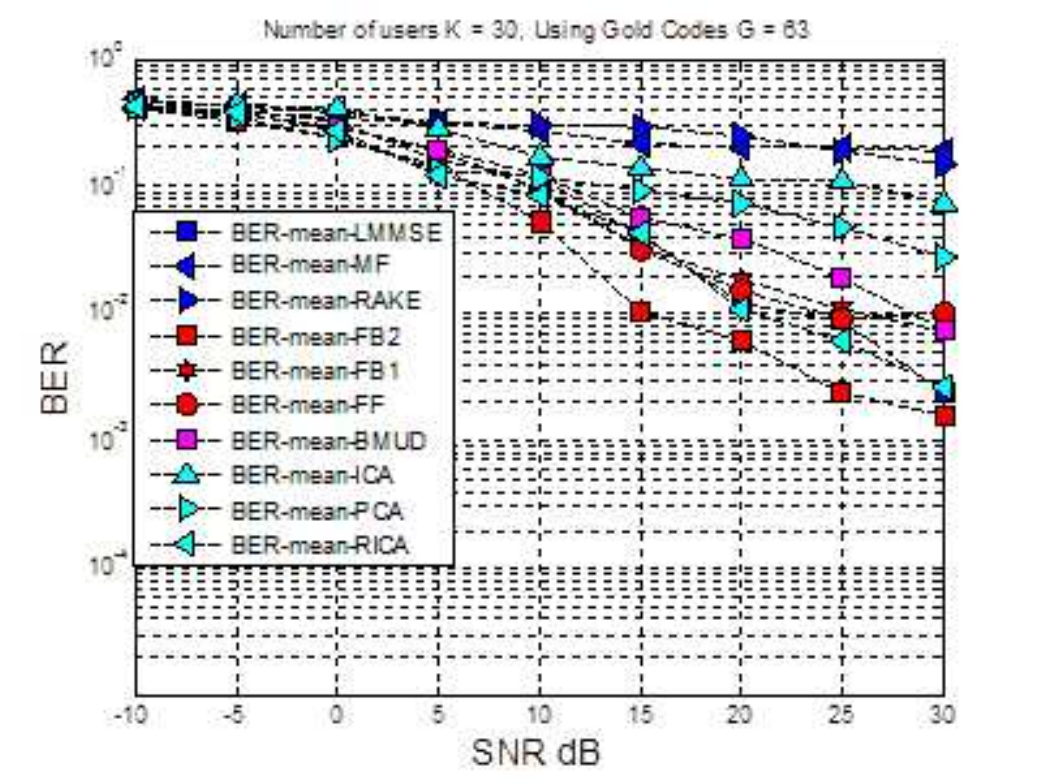}
}
\end {subfigure}

  \begin {subfigure} [{Using 50 users }]{
          \includegraphics[height=200pt,width=300pt,angle=0]{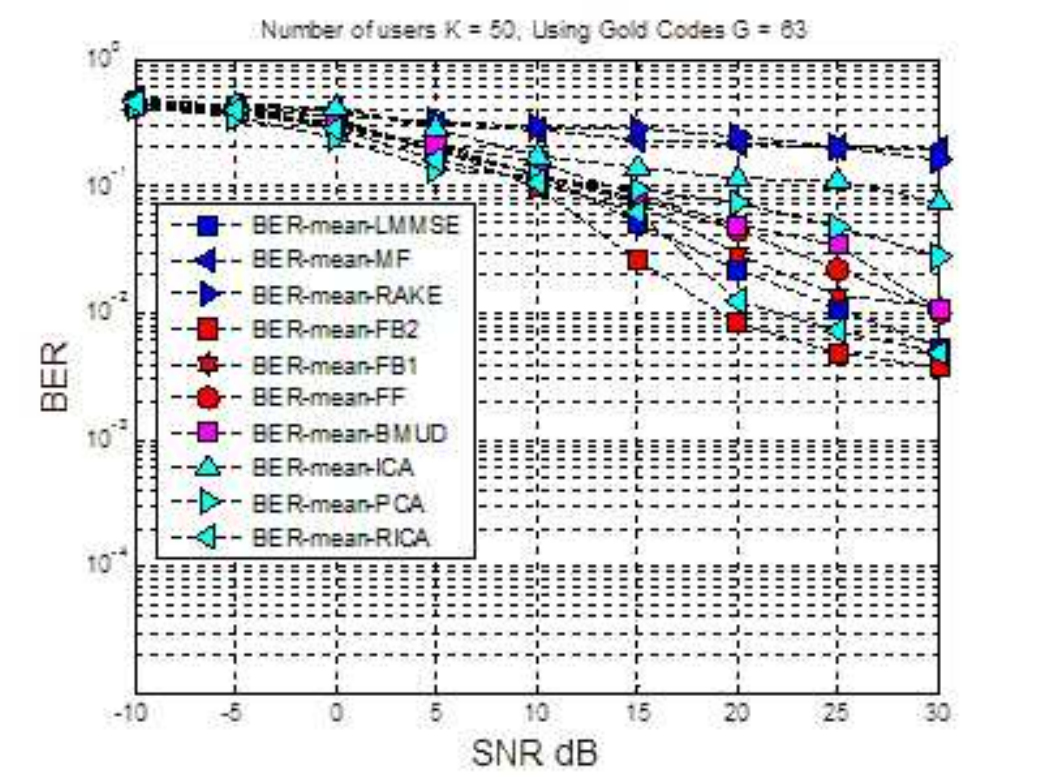}
}
\end {subfigure}
\caption{Average BER as a function of SNR for WCDMA downlink. Using Gold codes G=63. (a) Using 30 users (b) Using 50 users}
\label{fig:2}
\end{figure*}

\begin{figure*}[ht]
\centering
    \begin {subfigure} [{Using 30 users}] {
          \includegraphics[height=200pt,width=300pt,angle=0]{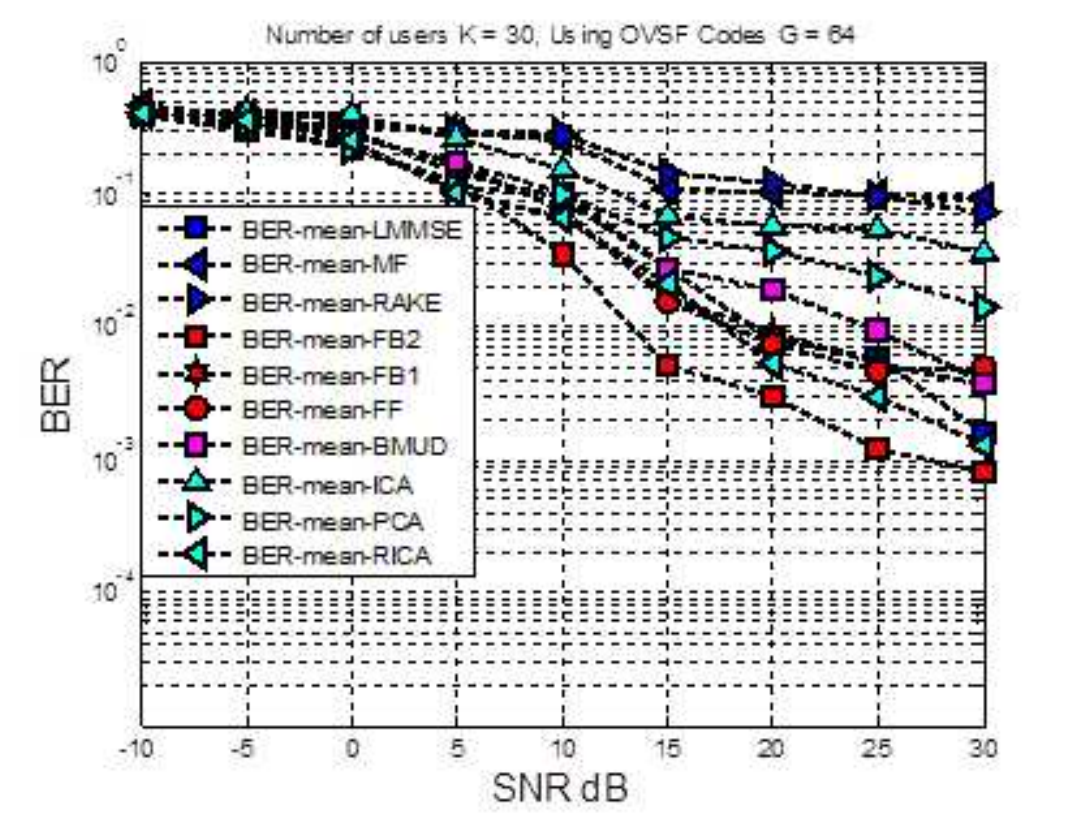}
}
\end {subfigure}

  \begin {subfigure} [{Using 50 users }]{
          \includegraphics[height=200pt,width=300pt,angle=0]{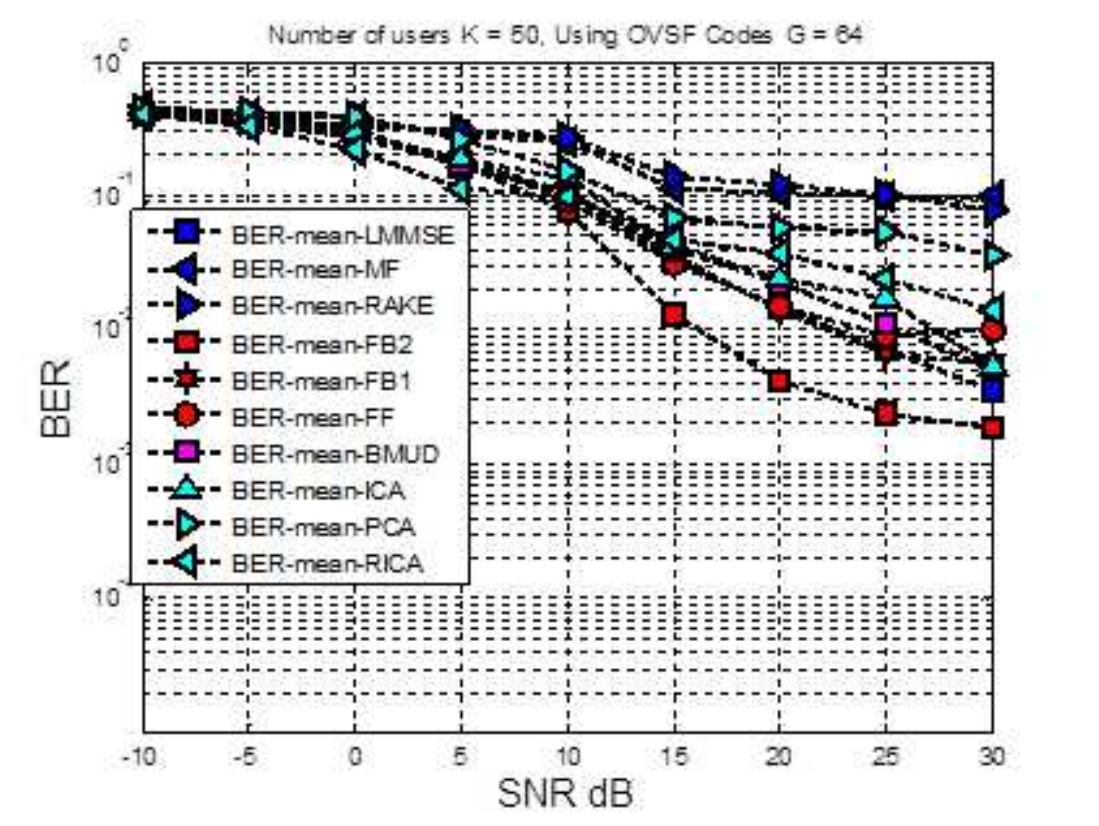}
}
\end {subfigure}
\caption{Average BER as a function of SNR for WCDMA downlink. Using OVSF codes G=64. (a) Using 30 users (b) Using 50 users}
\label{fig:2}
\end{figure*}

It is also worthwhile to compare the presented algorithms with a relatively large data sample set. Thus, Fig. 10 and Fig. 11 present the performance of the various detectors with fairly long sample set, namely,  M=30 000 in each of the DS-CDMA and WCDMA systems. It is noted that the benchmark LMMSE detector performs much better for high SNR. It is plausible to assume that the LMMSE detector becomes better than other detectors under good SNR conditions. However, the proposed algorithm based on FB-II has exceeded the LMMSE detector at all SNRs less than 22 dB. 

\begin{figure*}[ht]
\centering
    \begin {subfigure} [{Using Gold codes G=63}] {
          \includegraphics[height=200pt,width=300pt,angle=0]{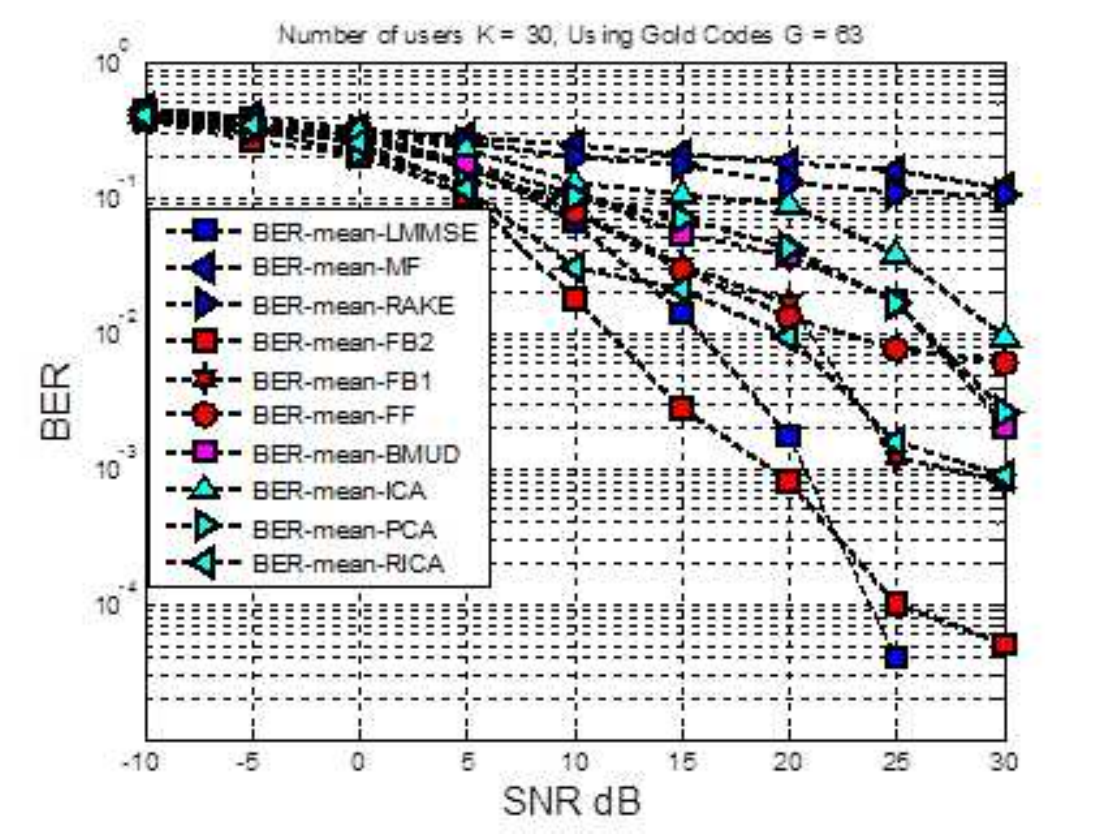}
}
\end {subfigure}

  \begin {subfigure} [{Using OVSF codes G=64}]{
          \includegraphics[height=200pt,width=300pt,angle=0]{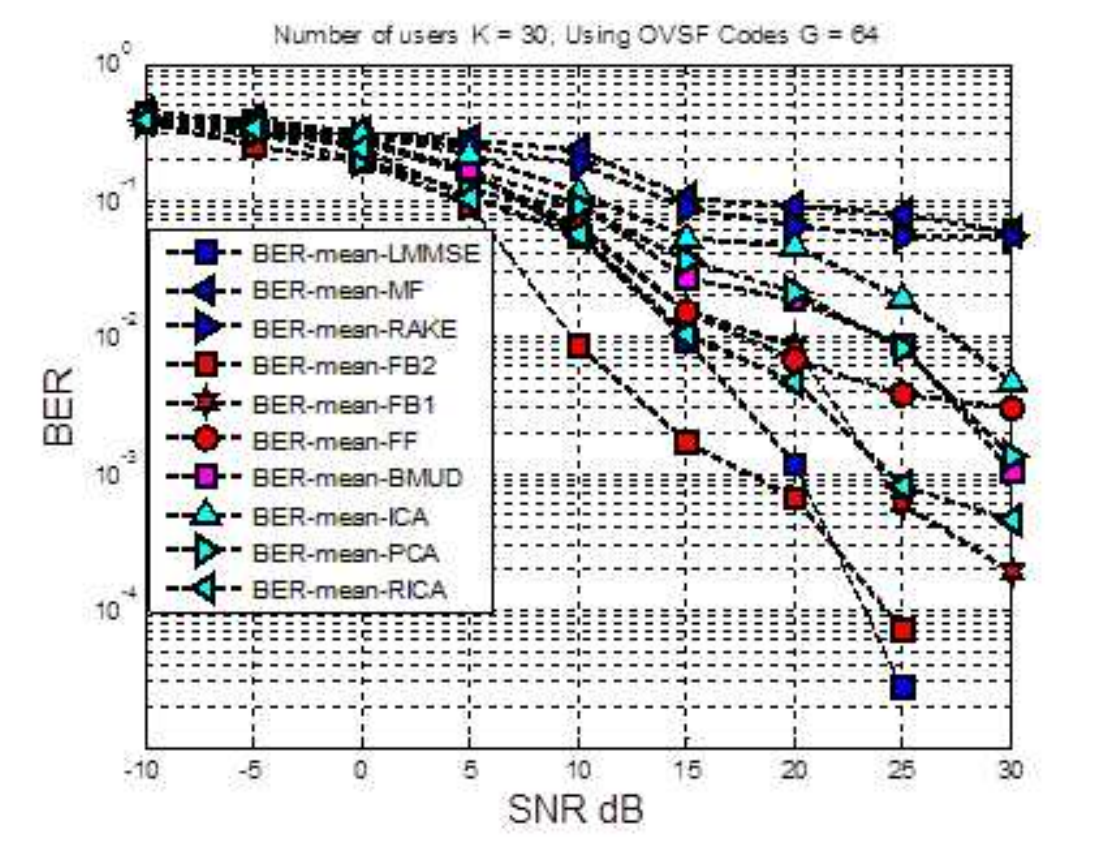}
}
\end {subfigure}
\caption{Average BER as a function of SNR for DS-CDMA downlink. For 30 users (a) Using Gold codes G=63. (b) Using OVSF codes G=64.}
\label{fig:2}
\end{figure*}

\begin{figure*}[ht]
\centering
    \begin {subfigure} [{Using Gold codes G=63}] {
          \includegraphics[height=200pt,width=300pt,angle=0]{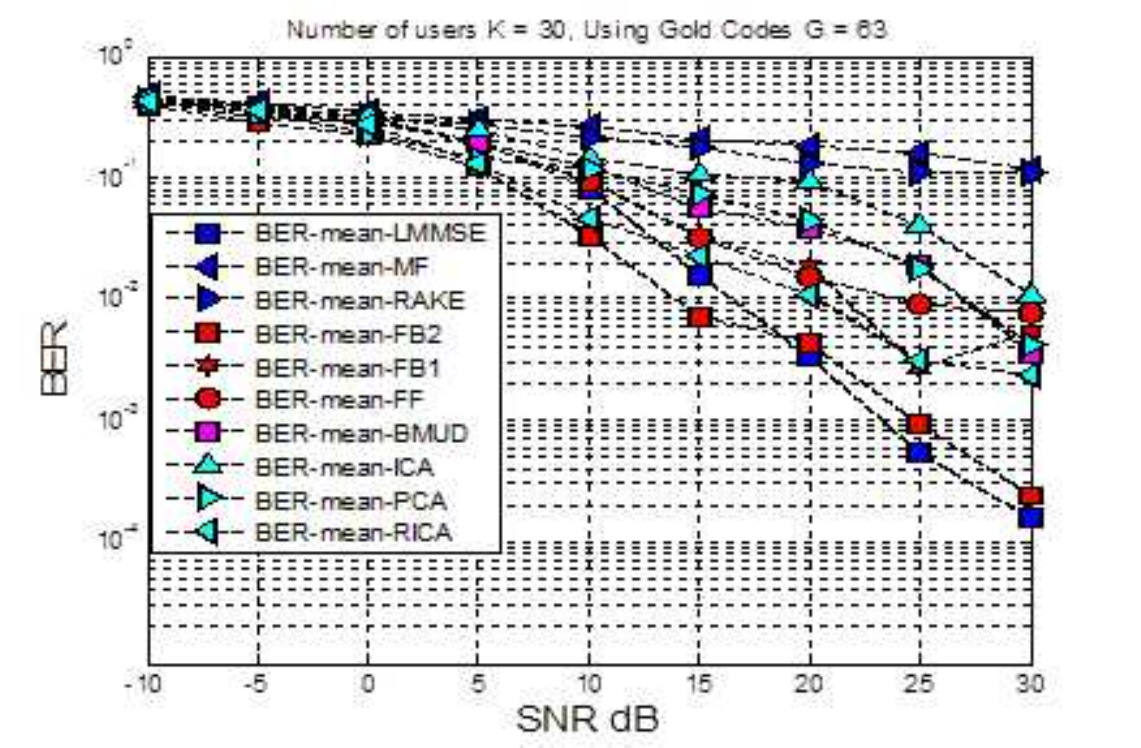}
}
\end {subfigure}

  \begin {subfigure} [{Using OVSF codes G=64}]{
          \includegraphics[height=200pt,width=300pt,angle=0]{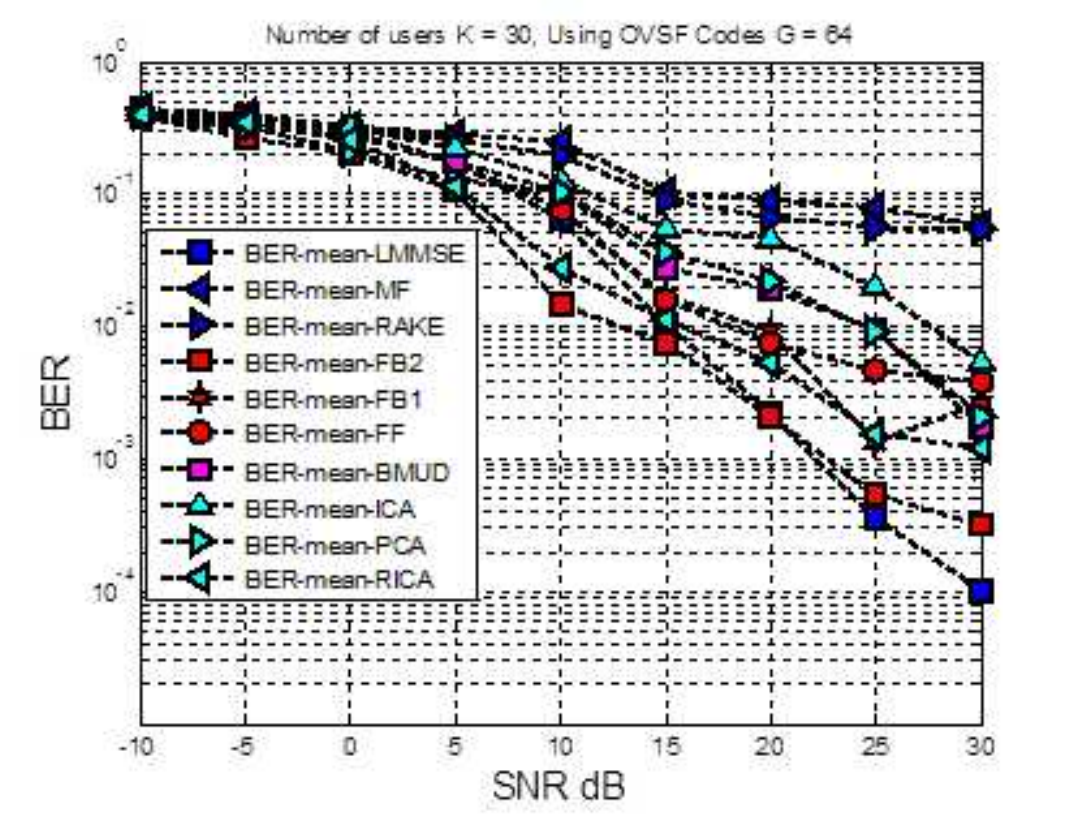}
}
\end {subfigure}
\caption{Average BER as a function of SNR for WCDMA downlink. For 30 users (a) Using Gold codes G=63. (b) Using OVSF codes G=64.}
\label{fig:2}
\end{figure*}

Finally, we evaluate the effect of the number of users and the size of the sample set on the performance of the proposed FB-II method in Figs. 12 and 13, respectively. In Fig. 12, the simulation results show the BER vs. SNR with various K users at 500 symbols for each user for blind multiuser detection based on the FB II detector. As expected, Fig. 12 shows that the FB-II detector decreases in performance as K, the number of users, is increased. Moreover, Fig. 13 shows the simulation results of BER vs. SNR with 30 users (K=30) for various data samples (M). The proposed FB-II algorithm appears robust and performs resonably well, and it is obvious that its performance improves more consistently as M increases by mitigating the MIA. 

Overall, the proposed variant detectors and algorithms perform well in solving the symbol estimation problem in the DS/WCDMA downlink system, especially when the size of the sample set is reasonably small.

\begin{figure}
 \includegraphics{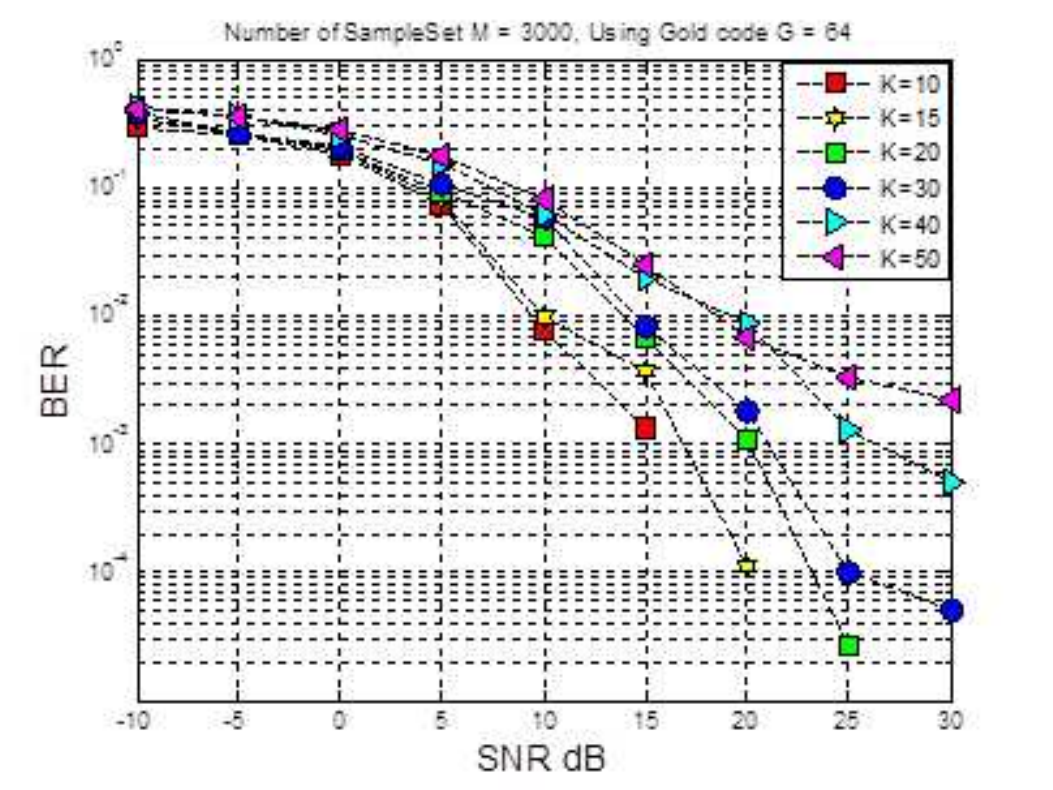}
\caption{Average BER as a function of SNR for various number of users  K}
\label{fig:1}       
\end{figure} 

\begin{figure}
 \includegraphics{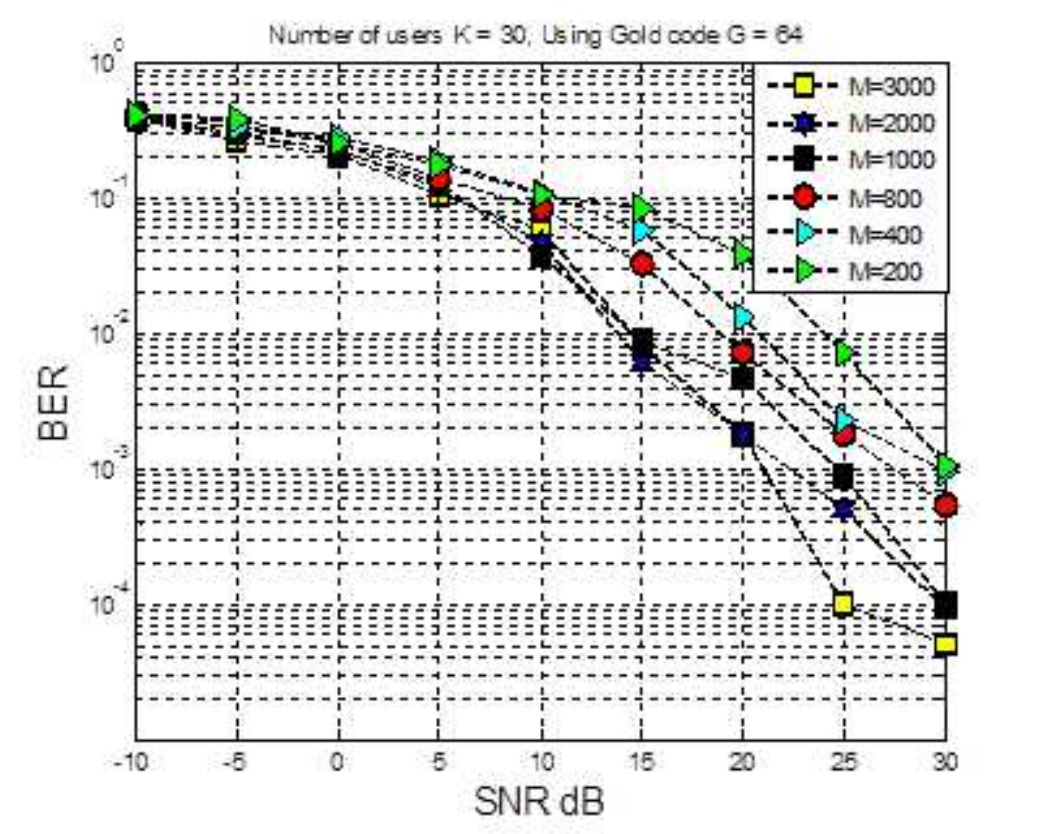}
\caption{Average BER as a function of SNR for various sample sets M}
\label{fig:1}       
\end{figure}

\section{CONCLUSION }
\label{sec:1}
We have  presented formulations, derivations, and subsequent extensive simulations of various filtering algorithms within various structures for multi-user detection in CDMA based systems. We have developed three blind multiuser detectors based on different filtering structures and three algorithms, namely, ICA, RICA and PCA. The results appear to show that the proposed structures perform well in the symbol estimation problem in DS/CDMA systems and more generally outperform all  other detectors, including the LMMSE detector. Our results also show that MAI can be mitigated by the proposed algorithms, particularly the proposed FB-II detector. Although the FB-II detector further improves as the size of the sample set increases, the results show that it performs well even when the sample sets are small-- unlike the LMMSE detector. Finally, the proposed algorithms, unlike the adaptive LMMSE detector, do not require the spreading codes of the interfering users. While these detectors are more suitable for the downlink case, they can also be used in the uplink case as well. 
\newline


\section*{Acknowledgement}
This research was supported in part by the NSF Grant ECCS-1549517.

%
%




\end{document}